\documentclass[5p,final]{elsarticle}
\usepackage{amsfonts}
\usepackage{amssymb}
\usepackage{amsmath}  
\usepackage{amsthm}
\usepackage{mathtools}
\usepackage{enumitem} 
\usepackage{geometry}
\usepackage{graphicx}
\usepackage{algorithm}
\usepackage{algcompatible}
\usepackage[colorlinks=true, allcolors=blue]{hyperref}
\usepackage{hyperref}
\usepackage[utf8]{inputenc} 
\usepackage[english]{babel} 
\usepackage{url}
\usepackage{wrapfig}
\usepackage{subfiles}
\usepackage{xfrac}
\usepackage{multirow}
\usepackage{cleveref}
\usepackage{balance}
\usepackage{booktabs}
\usepackage{tikz}
\usetikzlibrary{shapes.geometric, arrows}
\usepackage{tikz-3dplot}
\usepackage[squaren]{SIunits}
\usepackage{caption}
\usepackage[normalem]{ulem}
\usepackage{subcaption}
\usepackage[figuresleft]{rotating}
\usepackage{lipsum}
\usepackage{todonotes}
\usepackage{float}
\usepackage{pgfplots}
\pgfplotsset{compat=newest}

\usepackage{pgffor} 

\newcommand{\bc}{\begin{center}}
\newcommand{\ec}{\end{center}}
\newcommand{\be}{\begin{equation}}
\newcommand{\ee}{\end{equation}}
\newcommand{\bea}{\begin{eqnarray}}
\newcommand{\eea}{\end{eqnarray}}
\newcommand{\beq}{\begin{eqnarray*}}
\newcommand{\eeq}{\end{eqnarray*}}
\newcommand{\bv}{\left( \begin{array}{c} }
\newcommand{\ev}{\end{array} \right) }

\newcommand{\pvwap}{p_{_{\mathrm{VWAP}}}}

\makeatletter
\def\ps@pprintTitle{%
  \let\@oddhead\@empty
  \let\@evenhead\@empty
  \def\@oddfoot{\reset@font\hfil\thepage\hfil}
  \let\@evenfoot\@oddfoot
}

\usepackage{natbib}

\begin{document}   
\selectlanguage{english}
\title{Many learning agents interacting with an agent-based market model}
	\author[1]{Matthew Dicks}
	    \ead{matthew.dicks@alumni.ac.za}
	\author[1]{Andrew Paskaramoorthy}
	\ead{andrew.paskaramoorthy@uct.ac.za}
    \author[1]{Tim Gebbie}
    	\ead{tim.gebbie@uct.ac.za}
	\address[1]{Department of Statistical Sciences, University of Cape Town, Rondebosch 7701, South Africa}
	\date{\today}
    \begin{abstract} 
    We consider the dynamics and the interactions of multiple reinforcement learning optimal execution trading agents interacting with a reactive agent-based model of a financial market in event time for a single traded instrument. The model represents a market ecology with 3-trophic levels represented by: optimal execution learning agents, minimally intelligent liquidity takers, and fast electronic liquidity providers. The liquidity takers cannot learn to switch strategies. Learning is restricted to an optimal execution agent class that includes buying and selling agents. These agents can use combinations of limit orders and market orders. Their reward function explicitly balances trade execution slippage against the penalty of not executing the order timeously. This demonstrates how multiple competing learning agents impact a minimally intelligent market simulation as functions of the number of agents, the size of agents' initial orders, and the state spaces used for learning. Phase space plots are used to examine the dynamics of the various model configurations. Including learning agents in chartist-fundamentalist-noise models for a single traded instrument improves conformity with the empirical stylised facts but is insufficient to recover the complexity observed in empirical data.
    \end{abstract}
	\maketitle

\section{Introduction}

At high-frequency time scales, we can investigate how prices change at the level of individual trades revealing several remarkable micro-structural stylised facts \cite{pagan1996econometrics,hasbrouck1988trades,cont2001empirical,bouchaud2002statistical,bouchaud2003fluctuations,potters2003more,chakraborty2011market,cartea2015algorithmic}. The most important of these stylised facts include: the long-memory of order flows and absolute returns, the distribution of rare events, and the power-law of price impact. Taken together, these stylised facts motivate a model for price dynamics based on order flow and liquidity provision, arising from the strategic behaviour of different classes of heterogeneous agents, operating at different time scales and under asymmetric information. 

These agents conceal private information to prevent adverse selection, which limits the available liquidity and has given rise to the notion of latent supply and demand for liquidity not reflected in the visible order-book. These micro-structural stylised facts are plausibly thought to arise from the interaction of latent supply and demand with revealed liquidity, facilitated by optimal execution agents. However, by definition, data pertaining to latent demand and supply is not publicly available\,---\,it is hidden.

Agent-Based Models (ABMs) have been extensively used to explain subsets of various low-frequency stylised facts arising from the interactions of heterogeneous agents; most commonly chartists and fundamentalists within a minority game setting\cite{ChalletZhang1997,ChalletMarsiliZhang2000,Marsili2001}. However, some micro-structural stylised facts can be directly attributed to the behaviour of optimal execution agents and isolating these features has received far less attention within the ABM literature. The execution of any parent order arising from latent demand incurs trading costs in the form of price impact. Optimal execution agents will try to minimise this impact, which arises from limited liquidity and incomplete information, through the use of strategic order-splitting. 

Here, a large parent order is split into smaller child orders to be sequentially executed. Empirically, order-splitting appears to be widespread in real markets, and is used to explain the observed persistence of order flow and the long-memory of the size of returns. Furthermore, it is now well-established that realised price impact is concave with a power-law relationship, which has been attributed to the transfer of latent liquidity to the visible limit order book in response to a change in price \cite{Donieretal2015,Farmeretal2013}. In light of this, how should optimal execution agents be defined in order to minimise costs and simultaneously reproduce these stylised facts? 

Firstly, it appears that realistic execution agents must also be able to use limit orders (LOs) to execute parent orders, to facilitate the transfer of latent demand and supply to the limit order book, and hence recover the concave price impact function. Next, we can distinguish between static and dynamic execution strategies. Markets are not automatically efficient in the sense of price predictability, suggesting there is some form of t\^atonnement. 

Static strategies typically require market efficiency or asymmetric information access to be optimal, {\it e.g.} consistency with linear price impact \cite{Kyle1985,HubermanStanzl2004,Farmeretal2013}. In contrast, the observed concavity of price impact \cite{bouchaud2003fluctuations,gould2013limit}  provides both a normative and a positive argument for dynamic execution strategies that are aware of market conditions. Since a correct {\it a priori} specification of the data-generating process is unrealistic, it seems that some form of learning is necessary for dynamic strategies to be operational in practice. However, in the presence of many learning agents acting competitively, is learning still possible? This seems to depend on how market dynamics increase in complexity with the number of additional agents and the correlation between their payoffs \cite{galla2013complex, sanders2018prevalence, pangallo2019best}. 

In summary, the process that seems to dominate high-frequency phenomena appears to be the coupling between the low-frequency latent demand with the high-frequency mechanics, {\it i.e.} the market-microstructure of continuous-time double auction markets in the presence of limited liquidity. We speculate that the role of learning and the choice of order types in optimal execution are fundamental to this process.

To explore this, we simulate how latent demand is revealed in a single stock financial market using an ABM to capture the market environment into which we introduce both single and many optimal execution agents that engage in learning. The execution agents recieve their parent orders from agents generating the latent demand, and these agents each execute a single parent order in a market environment that consists of (minimally intelligent) chartists, fundamentalists, and high-frequency market makers. Model-free learning is incorporated using a simple (Multi-Agent) Reinforcement Learning specification. From these simulations, we examine how different model specifications affect stylised facts. 

We investigate whether the addition of the new agent-type\,---\,the execution agents with learning\,---\,to the traditional agent classes used to define the learning environment, can provide a more complete description of market ecology both in terms of stylised facts, but also by trying to recover the empirically measured market complexity. To do this, we view the ABMs as nonlinear dynamical systems and compare their complexity as measured by Grassberger-Procaccia correlation dimension plots (See Figure \ref{fig:embeddingdimensions}). 

However, our main contribution is to demonstrate how different model specifications affect stylised facts. This yields several key findings. Firstly, we find that {\it learning decreases the persistence in order flow}, with some evidence that learning can also decrease the memory in the absolute returns. Second, we find that {\it the persistence of order flow is largely determined by the difference in the number of buying and selling agents}. Third, we find that {\it increasing the number of agents increases the persistence of order flow}. Fourth, the ability to use LOs to execute a parent order results in lower price impact and faster decay in the size of absolute returns. This suggests that {\it a good approach to optimal execution will always be a judicious combination of limit-orders and market-orders}\footnote{Here limit-orders are resting orders passively inserted into the order-book with a price and volume requirement, while market orders are orders for immediate execution that demand a particular volume irrespective of the prevailing price.}. Surprisingly, we did not find conclusive evidence that learning reduced price impact. Lastly, the inclusion of many execution agents endowed with learning, trading a single stock, is not able to recover the complexity observed in real-world data.

The missing complexity demonstrates that our agent-based models are incomplete and we speculate this is  because of the exclusion of intra-order book network effects; that market models will be incomplete unless they include the cross-order book trading. The missing complexity is substantial. Modelling order-books in isolation from other order-books and strategic agents may well be a first-order effect and is probably the most important observation arising from this project even when the model itself does reasonably recover the appropriate dynamics.

The rest of this paper is organised as follows. In Section \ref{sec:background}, we briefly review the literature to motivate our investigation into optimal execution with learning. In Section \ref{sec:learning-agents}, we specify a novel learning agent that can post market orders (MOs) and limit orders (LOs) to execute a parent order, and examine its learning dynamics (section \ref{ssec:convergence}). The underlying model for the environment is that provided by \cite{Dicksetal2024ABM}. In section \ref{sec:results}, we present the remaining results of our study, which includes the analysis of stylised facts in Section \ref{ssec:stylisedfacts}, and an investigation into the market dynamics and complexity in Section \ref{ssec:complexity}. Finally, in Section \ref{sec:conclusion}, we conclude by summarising our study and indicating possible future directions of research. 

\section{Background and Motivation}\label{sec:background}

\subsection{Reinforcement Learning for Optimal Execution}

An important question in formal models of learning is determining what conditions allow for successful learning. Learning is said to be possible when, given a sufficient amount of data, the errors of the learner's outputs can be made arbitrarily small. When it is difficult to show analytically that an algorithm can learn, the ability to learn is intimated from measuring performance on test data. In sequential decision problems, asymptotic convergence guarantees have been derived for many reinforcement learning algorithms, such as Q-learning, under the assumption that the agent's environment is stationary \cite{littman1994markov}. Here, stationarity refers to state transition probabilities and the reward function. In financial markets, where the dynamics of the environment are changing, is learning possible?

Intuitively, we may guess that under a changing environment, a learning agent would have to continually relearn its policy to be optimal under the prevailing conditions. Thus, it would seem that asymptotic convergence may not be possible, but this need not be catastrophic to learning. Learning algorithms may still be able to outperform rule-based counterparts, but inferring performance on static data may be misleading. 

\citet{Dicksetal2024ABM} show that a simple learning agent could outperform a Time Weighted Average Price (TWAP) agent even without policy convergence. Here the learning dynamics where of a simple RL execution agent in an ABM where non-stationary dynamics in the environment are primarily driven by the interactions of minimally intelligent agents. However, the difference in policy over consecutive training periods did appeared to converge.

This work extends \cite{Dicksetal2024ABM} by studying the learning and market dynamics that arise from many learning agents interacting within an environment that includes minimally intelligent agents.
Interactions between minimally intelligent agents can produce nonlinear dynamics, but agents' decision rules do not change over time, limiting the variation in the environmental dynamics. However, in the presence of multiple learning agents, the best policy of an agent changes in response to the changing environmental dynamics, which in turn changes with the other agent's policies \cite{busoniu2008comprehensive}. Studying optimal execution in a MARL framework within an ABM can lead to a better understanding of how increasing the number of RL agents affects the agents' ability to learn, as well as how these affect the underlying ABM. 

In multi-agent reinforcement learning, an important problem is the appropriate specification of learning goals, which typically involves a trade-off between stability and adaption \cite{busoniu2008comprehensive}. The former is desirable for inferential purposes and establishing generalisation, but the latter is desirable for continual performance improvements. For example, learning stability can be established by specifying convergence to an optimal Nash equilibrium as a common learning goal. The combination of the two broad learning goals can be seen as having each agent converge to a stationary policy where each agent's policy is the best response to all the other agents' policies. Convergence to a Nash equilbria is easier to establish when agents are fully cooperative (agents have identical reward functions) or competitive (agents' rewards are zero-sum), in comparison to when agents are trained independently without a common goal \cite{bowling2000convergence}. In the latter case, asymptotic performance guarantees of (single-agent) Q-learning may fail \cite{tan1993multi}, but training agents independently still can work well in practice \cite{zhang2021multi, foerster2017stabilising}. 

\subsection{Learning in Complex Multi-player Games}

This work is further motivated by the study of learning dynamics in complex games \cite{galla2013complex, sanders2018prevalence, pangallo2019best}, where complexity is measured in terms of the number of possible actions. In particular, \citet{galla2013complex} considers a two-player game, and show how the learning dynamics of an RL algorithm called experience-weighted attraction (EWA) are affected by the correlation between the player's payoffs and the memory parameter of the learning algorithm. \citet{galla2013complex} finds that the learning dynamics of players' strategies can be separated into three different regimes. Namely, \textit{i)} strategies can converge to a unique fixed point if agents have short memory, and the correlation between the players' payoffs becomes increasingly negative (ie. competitive), or \textit{ii)} if correlations are negative but players have long memory, then learning dynamics can be chaotic or converge to limit cycles, and \textit{iii)} if players have long memory and payoffs are positively correlated (ie. cooperative), then learning has a multiplicity of fixed points. In the second regime in particular, strategies are essentially random and learning is not possible. In two-player normal-form games, a possible reason for this is that as games get more complicated and/or more competitive this causes best reply cycles to become dominant, which \cite{pangallo2019best} shows to be a good predictor of the non-convergence of several learning algorithms. In games involving more than two players, \citet{sanders2018prevalence} show that the parameter range in which learning converges to a fixed point becomes smaller as the number players increases. This implies that chaotic behaviour is characteristic of many-player games. 

In summary, as games become more complex, and agents payoffs become increasingly independent, and the number of players increases, equilibrium becomes more unlikely. So, if normal-form games are a good model for the games market participants play, and if these participants can be modelled reasonably well by learning algorithms, this body of work calls into question the common assumption of equilibrium in economics and finance. However, a key constraint related to the carrying capacity of real-world markets is that of liquidity: without sufficient liquidity, trading decisions can not be made, compromising both the ability to learn and the profitability of the corresponding policy. Hence, to understand how the scaling insights of many player games can impact financial markets, one needs to include a broader ecosystem that captures these salient constraints: liquidity and the indirect cost of trading. 

\subsection{Market Ecology}

The interactions of market participants can be viewed from an ecological perspective, forming a useful conceptual framework to develop ABMs to investigate disequilibrium price dynamics. Agent classes are specified in terms of strategies, analagous to ``species" in ecological systems, which are rules describing how agents make trading decisions and hence govern their interactions \cite{farmer2002market}. Whilst market ecology is far from a formal theory of market functioning, two classification systems seem to have emerged in the literature: i.) Chartist-Fundamentalist-Noise (CFN) models and ii.) Liquidity-Provider (LP) and Liquidity-Taker (LT) models \cite{bouchaud2009markets}. In both classification systems, agent interactions are intermediated by price. Many of these types of models allow strategy switching \cite{DAY1990299,BROCK19981235, ChairellaDieciHe2009} as a form of adaption and naive learning.

There is a key nuance relating to where and how learning can operationally take place in real financial markets with these types of agent representations. In real financial markets, traders and money managers do not easily change their trading strategies because of the regulatory environment in which they reside {\it e.g.} risk capital allocations on trading desks are made based on strategy definitions and trading simulations, while asset management mandates to money managers are most usually stated {\it a priori} as part of agreements between providers of capital and those trading and investing, even in CTA funds. For this reason, we favour a hierarchical approach where there is strategy die-out and the emergence of new traders with particular {\it a priori} strategies for the design of the fundamentalists and chartists. This is part of the key design decisions made in this work and in prior work that is used to define the learning environment \cite{Dicksetal2024ABM}. 

It should be noted that some of the early prior literature bundle learning feedbacks and strategy definitions into a single model layer and thus risk conflating the impact of learning and adaption with the impact of die-out and feedback without learning (by allowing in-layer strategy switching) \cite{BROCK19981235, ChiarellaHe2005}. Similar care is prudent when considering the emergence of herd behaviours \cite{DAY1990299, Lux1995} without carefully considering the driving force behind herding itself, which is most often driven by higher order or adaptive behaviour of predators. In our work, the reinforcement learning component relates entirely to optimal execution in the presence of low-order strategies that are either fundamentalist (and contrarian) or chartists (and trend following) and learning-based adaption is separated into an isolated model structure. In part, this is why in this work we place some weight on the descriptive narrative surrounding the trophic level structure. These structures are incredibly important in real ecosystems for system stability for the same reason that we  expect them to be important in financial markets \cite{AulettaEllisJaeger2008}.

The choice we have made here is to follow the approach of \citet{WilcoxGebbie2015} where it was argued that what has been learnt about more general complex systems \cite{AulettaEllisJaeger2008,Ellis2008,Ellis2012} should probably be considered, and transferred into how models are built in financial market ecosystems. Particularly if the most important feature of real stable complex systems is a hierarchy of causal complexity \cite{AulettaEllisJaeger2008}\,---\,that the causal classes create system robustness through network effects that are not entirely bottom-up. 

In our model setting, the reinforcement learning agent would be an example of Top-Down Causation class IV (TDC4) engaging in feedback control with
adaptive goals \cite{Ellis2008,Ellis2012,WilcoxGebbie2015} while the lower level in which the fundamentalist and chartist reside are Top-Down Causation class II (TDC2) as agents engaging in non-adaptive information control. It should be noted that the market-makers facilitating liquidity provision in actual markets would have an objective function and engage in risk management, or rather money management portfolio control, and would appear to be more like the Top-Down Causation Class III (TDC3) agents engaging in adaptive selection. Our implementation is different because we feel that we can correctly capture the high-frequency traders being mechanistic liquidity providers at first order, and as such, in our implementation appear more like a TDC2 actor, than a TDC3 actor. 

The reason for this is that we have excluded intra-order book traders and the inventory management component for real risk traders and market makers. Here, we have chosen the simplest model that we felt captures the model design narrative we are interested in while separating learning as a higher-order function separate from the niche being created that learning can exploit (the environment created by the fundamentalists and chartists in the presence of liquidity providers). We now consider the learning environment more carefully \cite{Dicksetal2024ABM}.   

Chartists determine trading behaviour purely on past performance, whereas fundamentalists determine trading behaviour based on current price relative to some subjective valuation. Chartists and fundamentalists have opposing effects on price dynamics, and their interaction is able to produce many stylised facts observed in intra-day and lower frequencies (mesoscale), including clustered volatility and some additional fattening of the tails in returns. In contrast, the liquidity provider and liquidity-taker classification is useful to explain additional microstructural stylised facts in terms of how information is processed into prices by agents operating at different timescales, whilst keeping prices unpredictable {\it e.g} some order-flow memory. 

In particular, liquidity-takers operating at low frequencies (particularly the fundamentalists) with large liquidity demands, can only trade incrementally owing to the limited available liquidity provided by higher frequency market makers. In between the large low-frequency liquidity takers and high-frequency market markers, are chartists acting at shorter time intervals acting on information contained in price changes. An additional feature of our model formulation is that we consider the fundamentalists and chartists to be minimally intelligent in the sense that the price levels that they trade relative to are randomly generated, and they are operationally trading noise in an environment with limited liquidity. From trading the noise, structure emerges that the learning agents can respond to.   

To capture an ecology with limited liquidity and reasonably realistic market impact, we postulate a model in terms of 3-trophic levels defined respectively by three agent classes: learning agents, liquidity takers, and liquidity providers. In any given epoch, the liquidity takers can have increasing and decreasing profits and face gamblers ruin, but they cannot switch strategies. This is similar to the thinking used in a 3-trophic level model of carnivore-herbivore-plant systems. Here, we have learning agents as the ``carnivores", or predators that engage in opportunistic behaviours which necessarily require adaption and hence, learning. We have the liquidity takers, both the fundamentalists and the trend-followers, as ``foraging herbivores" since their activity is rule-based and hence ``passive" in their consumption of liquidity. Liquidity is the ``food" provided by ``plants", which are uninformed High-Frequency Traders (HFTs), or rather Electronic Liquidity Providers (ELPs). This suggests not only a financial market equivalent of a ``food web" but also the inherent nonlinearity and feedbacks. Again, we include this because we believe that financial market model builders can learn from other domains that have become adept at model building and testing in complex systems\,---\,in particular, those working in Ecosystem model building and simulation.   

How trophic levels are defined in financial markets is necessarily contested, and can be fluid as markets change and adapt because financial markets are reflexive \cite{Polakowetal2023}. However, this metaphor frames an the narrative around the role of learning agents within our financial market ecosystem\,---\,specifically the nature of strategic order-splitting in an environment with a highly constrained carrying capacity. The point of using such a narrative this way is to try to move the model building away from mechanistically implementing this or that convenient mathematical feature, and then coming up with a narrative that uses the mathematical trick, and then testing against the stylised facts; to probing how we think about the agents and the environment itself that the model aims to represent. Here, we aspire to create a link between the narrative, the models, and the observations on an equal footing.  

We apparently do not need learning for the first two trophic levels of our system since static mechanistic rule-based responses to the market states seem to be sufficient to recover almost all of the necessary stylised facts of the environment \cite{Dicksetal2024ABM}. Learning seems to only become important as a coupling mechanism between the visible market (the learning environment) and the latent order book. This provides a mechanism to drive strategic order splitting that can be used to fine-tune the observed balance between order-flow persistence, price volatility and clustering, and the observed extreme events in the complete market. This is particularly interesting because we know from the prior work that further tuning the parameters describing the environment cannot do so \cite{Dicksetal2024ABM}.

Whilst there are feedbacks between each class of agents in the different trophic levels, our framework suggests a hierarchy based on dependency. Although not a functioning market, the presence of ELPs is sufficient for the existence of a persistent order-book, hence providing an environment for the other agent classes, and thus existing at the bottom of the hierarchy. In contrast, learning is situated in higher trophic levels \cite{AulettaEllisJaeger2008}, because the higher levels necessarily include adaption via feedback control\footnote{\citet{AulettaEllisJaeger2008} use the term {\it information control} to make clear the importance of: i.) information selection, and ii.) feedback control. We draw the interested reader to their Figure 7 \cite{AulettaEllisJaeger2008}. The key idea is that top-down causation by information control occurs when an equivalence class is established, and the information selection defining the operations of the class is conserved because of modularity, despite the variability of lower-level variables. This is part of why learning and adaption are typically found in higher levels, and not lower levels, in real-world systems.}. However, from simulation work and the sensitivity analysis we know that learning agents cannot adapt to ELPs alone because that would merely create noise -- this should be explored more carefully in future work and is outside of the scope of this project. This indicates the need for minimally intelligent agents at an intermediary level whose actions create learning opportunities by using the noise signal to generate exploitable but transient structure via emergence. This picture, in turn, would suggest that a market made entirely of learning agents would not be feasible because of the extremely limited liquidity and that in simulation the algorithms synchronise and via tight-coupling lead to system failure -- this should be explored more carefully in future work.

Concretely, we find that if only learning agents interact with noise traders learning is not possible unless the noise is synchronous across agents. We speculate that this implies that the carrying capacity of a market is a crucial feature. Any model that simulates the herding of learning agents presupposes sufficient liquidity to profitably support trading as well as learning and that the underlying data-generating processes are sufficiently synchronised to trigger coordination. Here the carrying capacity of the environment would be too low to support both the cost of learning, and then using it successfully, because other agents adapted similarly to the single learning agent with insufficient liquidity to take the necessary learning actions. This, in turn, points to the importance of the liquidity multiplier effect introduced by the minimally intelligent liquidity takers and potentially why a hierarchy of causality may well be a crucial model feature in realistic financial market models and simulations. 

\subsection{Limitations}

One question surrounding the study of ABMs is determining whether simulated paths represent general agent and market behaviour or are unique to modelling choices. In our implementation, we address this by separating various sources of variation and important model features in the modelling process. From the modelling side, we have ensured that learning only takes place at the highest level in the hierarchy of complexity in our model and that there is no strategic adaption in the middle level where our minimally intelligent liquidity takers reside. At the lowest level, we have zero-intelligence noise traders providing liquidity.

We retain the three-step approach for the development and calibration of the learning environment \cite{arxiv2021abm,Dicksetal2024ABM}: 1. sensitivity analysis, 2. calibration with a minimal set of known and re-used seeds using the relationships between model and parameter variations found from the sensitivity analysis, and 3. simulation using the same restricted number of seeds using the calibrated parameters for the training environment. 

The number of random seeds is chosen to crudely optimise the compute times while providing relative stability of the parameters and a learning environment, to faithfully capture the empirically measured environment. Many sample paths are simulated from each random seed to facilitate learning since including learning within an agent-based model imposes computational considerations that are at least an order of magnitude more onerous than a typical sensitivity analysis.

One of the drawbacks of ABMs is that they are computationally expensive to run, necessarily forcing pragmatic choices on the modeller. In this work, key sub-cases of the agent-based model were prudently chosen to explore the impact of both model and sample variations, whilst trying to prevent model and path variations that could confound our results. Since generating all possible combinations of parameters and paths is not computationally tractable, a subset must be pre-selected for simulation, which substantially restricts this type of simulation work. 

Consequently, it may appear there is the possible danger of selective bias in the presented results, where possible sub-cases, parameters, and paths are chosen {\it a-priori} to support a particular conclusion. However, owing to the nonlinearity of the system and the inclusion of various sources of noise, we do not know with any certainty the possible results corresponding to different parameter combinations, precluding the possibility of selective bias. However, conclusions from restricted results are never as robust as one would like, particularly when critiqued through the lens of linear statistical modelling techniques aimed at inference. Rather, our results are indicative and aimed to guide incremental model refinement to explain specific model features.  One key lesson is that, although we can capture most of the key stylised facts, we are not able to capture the observed market complexity, an experimental outcome that we think is very important; however, many of the narratives that emerge do make physical (or rather financial) sense. Because of the missing complexity we are very cautious about further model refinement and conclusions.

\section{Agent Specification and Learning}\label{sec:learning-agents}

We investigate the interaction of different types of optimal execution agents within a pre-calibrated event-based minimally intelligent ABM, which forms a training environment. The minimally intelligent agents consist of chartists, fundamentalists, and liquidity providers. The first two act as minimally intelligent agents without strategy switching, the last class is a zero-intelligence mechanistic noise trader. Their full specification, as well as details related to their learning dynamics, can be found in \citet{Dicksetal2024ABM}, which we do not restate here for the sake brevity. We study different cases, described in Table \ref{tab:RLagentscombinations}, where each case is defined by the number, type, and side ({\it i.e.} buying or selling) of the agents, to determine how each of these characteristics affects the stylised facts. 

The state space used for both agent classes corresponds to the smaller state space from \citet{Dicksetal2024ABM}, which is described in Table \ref{tab:discrete-state-space}. Consequently, the results of this study are directly comparable with the results of prior work as part of the model design. Similarly, the agent actions for MOs as based on multiples of a TWAP strategy to provide an additional dimension to compare with prior work. As such, this study is a direct extension of \citet{Dicksetal2024ABM}.

\begin{table*}[t]
        \caption{The best bid/ask volume ($v_{k}$) states are given, and the spread ($s_{k}$) states. Here the all provide the approximate probability of being in each states are based on the observed data at event $k$ and the simulated historical distributions. Explain how the bid and ask volume states works. Let $\tau = T_{0}/n_{_T}$, where $T_{0}$ is the total trading time for the RL agent and $n_{_T}$ is the number of time states. Let $i = X_{0}/n_{_V}$, where $X_{0}$ is the total amount of inventory to be traded and $n_{_V}$ is the total number of volume states. The actions associated with these states are visualised in Table \ref{fig:case3-policy-plot} and Table \ref{fig:case4-policy-plot}.}
        \label{tab:discrete-state-space}
        \centering
        \begin{tabular}{|c|rlc|rlc|rlc|rlc|}
            \hline
                 & \multicolumn{3}{c|}{Inventory states} &  \multicolumn{3}{c|}{Time states} &  \multicolumn{3}{c|}{Volume states} &  \multicolumn{3}{c|}{Spread states} \\
            State & \multicolumn{2}{c}{Range} & Prob. & \multicolumn{2}{c}{Range} & Prob.  & \multicolumn{2}{c}{Range} & Prob.  & \multicolumn{2}{c}{Range} & Prob. \\
             & \multicolumn{2}{c}{$ \cdot < x_{k} \leq \cdot$} & & \multicolumn{2}{c}{$ \cdot < t_{k} \leq \cdot$} & & \multicolumn{2}{c}{$ \cdot < v_{k} \leq \cdot$} &  & \multicolumn{2}{c}{$ \cdot < s_{k} \leq \cdot$} &  \\
            \hline
            1 & 0 & i & 0.2 & 0 & $\tau$ & 0.2 & 31   &         & 0.2 & 0 & 1       & 0.6     \\
            2 & i & 2i & 0.2 & $\tau$ & 2$\tau$ & 0.2 & 31   & 266     & 0.2 & 1 & 2       & 0.123   \\
            3 & 2i & 3i & 0.2 &  2$\tau$ & 3$\tau$ & 0.2 & 266  & 1453    & 0.2 & 2 & 3       & 0.062   \\
            4 & 3i & 4i & 0.2 & 3$\tau$ & 4$\tau$ & 0.2 & 1453 & 5209    & 0.2 & 3 & 7       & 0.091   \\
            5 & 4i & & 0.2 & 4$\tau$  & & 0.2 & 5209 & $\cdot$ & 0.2 & 7 & $\cdot$ & 0.124   \\
            \hline
            \end{tabular}
    \end{table*}

\subsection{Actions}

Following \citet{Dicksetal2024ABM} the volume of an outstanding parent order is $X_0$ and an order will have a volume trajectory $\{x_i \}_{i=1}^N$ so that $X_0=\sum_{i=1}^N x_i$. Then the $i^{th}$ decision point in a trading schedule gives a trade with volume $x_i$. Here all the agents use trading schedules that are multiples of a TWAP strategy for the original parent order. At each trade decision point there is an action $a_i$. For market orders, the action is a multiplier such that the submitted order has volume $a_i x_i$ from the volume trajectory that defines the TWAP trading schedule. For limit orders, the actions are the placement depth $a_{\delta}$ and rate of trading $a_{\nu}$. This suggests three basic types of execution agents: the benchmark TWAP trading schedule, a market order only trading schedule, and a trading schedule that combines market orders with limit orders. These three agent types are defined in Table \ref{tab:RL-specification-table}.   

The first type of agent is denoted ``Type S" and is a minimally intelligent execution agent characterised by a fixed TWAP schedule consisting of MOs only. The minimally intelligent type of execution agent is a benchmark agent and should be differentiated from the two learning agents.

We denote the first class of learning agents as ``Type I" agents \cite{Dicksetal2024ABM}. Type I agents have explicit order-splitting but only use MOs to interact with the market. Example actions as they relate to states for Type I agents are given in the heat map legend on the left of Figure \ref{fig:case3-policy-plot-A} and Figure \ref{fig:case3-policy-plot-B}. This means that Type I agents cannot directly interact with each other in the model but are inter-mediated only by the liquidity providers. 

Furthermore, in this study, we introduce a new class of agents, denoted ``Type II" agents, which can trade with both MOs and LOs. However, Type II agents can interact directly with each other and other agents as they use both LOs and MOs to trade. Example actions as they relate to states for Type II agents are given in the heat map legend on the left in Figure \ref{fig:case4-policy-plot}.

Both type I and type II agents use trading schedules that are multiples of TWAP and order sizes following \citet{Dicksetal2024ABM}. The action set of the Type I agents are given by $a_X=[0, 0.25, 0.5, 0.75, 1, 1.25, 1.5, 1.75, 2]$ giving a grid of actions from small orders to large orders as multiples of the TWAP strategy. For the Type II agents, MOs are placed into the market at some rate of trading $\nu$ as a function of machine time, while the LOs will be placed into the order-book at a depth $\delta$ from the mid-price. The order sizes will be as before. Here the actions will be the placement depth at an activation is $a_{\delta} = [0.01,1]$ (shallow or deep) and the rate of trading $a_{\nu} = \sfrac{1}{T}[100,10,1]$ (fast, moderate or slow) as a multiple of the total session length. 

\subsection{Rewards}

As in prior work \cite{hendricks2014reinforcement,Dicksetal2024ABM}, we considered using Perold's implementation shortfall \cite{perold1988implementation}, where the buying agent would try minimise the difference between the Volume Weighted Average Price (VWAP) achieved using its strategy and the hypothetical VWAP it would receive if it traded the entire parent order at the initial price, with no price impact using an Immediate Execution (IE) strategy. However, in this case, the minimum implementation shortfall is negative, which can incentivise buying at higher prices. Conversely, if you aim to maximise the implementation shortfall, to try to incentivise buying at lower prices, then you have the same problem arising from minimising the total cost because the agent learns not to trade. 

For the Type II agent, we specify a reward function that is symmetric between buying and selling agents, without resulting in agents learning degenerate policies. It is important to note that the simple reward function of the agent specified in \citet{Dicksetal2024ABM} is suitable for profit maximisation on the sell-side only. Attempting to construct a symmetric reward function applicable to buying agents through cost minimisation results in a degenerate policy which minimises trading costs by not trading. Thus, some form of penalty needs to be included in a suitable reward function to encourage trading. This is summarised in Table \ref{tab:RL-specification-table}.

\begin{table*}[th!]
\centering
\caption{Agents using strategic order splitting. Those engaged in learning their respective rewards and actions are given with the state-space size on $(n_t,n_I,n_S,n_v)$. All the agents below engage in learning except for the minimally intelligent agent $S$}
\label{tab:RL-specification-table}
\begin{tabular}{llccccc}
\toprule
Agent type & Reward & State-space & \multicolumn{3}{c}{Actions} \\ 
& &  Size & $\pm$ & Parent order & Order-type \\
\toprule
\citet{hendricks2014reinforcement} & short-fall & 5 & Sell & multiple of AC & MO's \\
\citet{Dicksetal2024ABM} & trading profit & 5,10 & Sell & multiple of TWAP &  MO's\\
\hline
Type S$^\pm$ & - & - & Buy/Sell & TWAP & MO's \\
Type I$^{\pm}$ & slippage - penalty & 5 & Buy/Sell & multiple of TWAP& MO's \\
Type II$^\pm$ & slippage - penalty &  5 & Buy/Sell & multiple of TWAP & MO's \& LO's \\
\bottomrule
\end{tabular}
\end{table*}

The aim is to formulate a reward function that is intuitive and minimises the cost on the buy side, maximises the profit on the sell side, and creates an incentive to trading. This can be achieved by a combination of the trader's slippage and a penalty for not trading. 

We start with $\pvwap(\cal X)$, the VWAP price received from the set ${\cal X}= \{x\}_{i=1}^n$ of all trades including the $n^{th}$ trade submitted by the $\ell^{th}$ RL agent.  The VWAP price found from all trades, excluding all the trades submitted by the $\ell^\mathrm{th}$ RL agent, is $\pvwap({\cal X} \setminus {\cal X}_{\ell})$. The difference between these two quantities measures the slippage between a given RL agent's trades, and the trades made by the rest of the market (potentially including other RL agents). 

The total time past in the simulation is $t$, the remaining inventory is $x_{\ell,n}$, and the total volume matched for the $n^\mathrm{th}$ order is $v_{n}$. We propose a penalty that increases exponentially as time increases, forcing the agents to learn to trade before the end of the simulation. The penalty is proportional to the inventory remaining and inversely proportional to the amount of volume matched for the last trade, ensuring that the penalty increases as the remaining inventory increases, and decreases as the amount traded by the $n_{th}$ order increases. The agent will want to trade more when there is a lot of inventory remaining. The penalty has two parameters: $\lambda_{r}$ controls how much effect the penalty term has on the reward function, and $\gamma_{r}$ controls the sensitivity to time.

Now, for the $n^{th}$ order in an episode the {\it reward function} for the $\ell^\mathrm{th}$ learning agent can be given as combination of the slippage and penalty: 
\begin{align}
    R_{\ell,n} = \underbrace{
        \pm \ln\left({
        \frac{\pvwap(\cal X)}{\pvwap({\cal X} \setminus {\cal X}_{\ell})}
        }\right)}_{\mathit{slippage}
        } 
        -  \underbrace{\left(\frac{x_{\ell,n}}{v_{n}}\right) \lambda_{r}  e^{{\gamma_{r} t}}}_{\mathit{penalty}
        }.
\end{align}
The first term is the {\it slippage}, and is positive for selling agents and negative for buying agents. By maximising this reward, we aim to get higher prices when we sell, and lower prices when we buy. The second term is the {\it penalty} term for not trading, which creates an incentive to trade, whether on the buy or sell side. 
 
Together, the slippage and penalty mean that the agents aim to minimise slippage and get the best prices relative to the rest of the market while being incentivised to trade. The return function for the $\ell^{th}$ RL agent is the accumulated reward: $\sum_{n} R_{\ell,n}$. This will be maximised by all the RL agents.

\begin{table}[h!]
    \centering
    \begin{tabular}{llll}
    \toprule
    Case \# & Types & \#Agents & Parent order size \\
      \hline
      0   &  - & 0 & - \\
      1   &  S$^-$ & 1  & 6\% ADV \\
      2   &  5S$^+$& 5  & $\frac{1}{5}$(6 \% ADV) \\
      3   &  I$^+$  & 1 & 6\% ADV \\
      4   &  I$^-$  & 1 & 6\% ADV \\
      5   &  I$^+$,I$^-$ & 2 & 3 \% ADV \\
      6   &  II$^+$ & 1 & 6 \% ADV \\
      7   &  II$^+$,II$^-$ & 2 & 3 \% ADV \\
      8   &  II$^+$,I$^-$ & 2  &  3 \% ADV \\
      9   &  5I$^-$ & 5 & $\frac{1}{5}$(6\% ADV) \\
      10   &  5II$^+$ & 5 & $\frac{1}{5}$(6\% ADV) \\
      11   &  5I$^+$, 5I$^-$ & 10 & $\frac{1}{10}$ (6\% ADV) \\
      12  &  5II$^+$, 5II$^-$ & 10 & $\frac{1}{10}$ (6\% ADV) \\
         \bottomrule
    \end{tabular}
        \caption{Different combinations of agents engaging in strategic order splitting. The first case is the agent-based model without any learning agents -- type 0. The minimally intelligent agents engaging in naive order splitting using a TWAP strategy are denoted as agents of type S. The agent types are either an acquisition agent ($+$) or a liquidation agent ($-$). The number of acquisition agents and liquidation agents is given for each case. 
        The overall order size is in multiples of $X_0 = 6\%$ of Average Daily Volume (ADV) for all agent classes to ensure that the market has similar liquidity across all the cases considered. The learning agents are then either of class I or II. Here class I only use market-orders and class II use both market-orders and limit-orders (see Table \ref{tab:RL-specification-table}). The state-space is of size 5 for all the learning agents.}
    \label{tab:RLagentscombinations}
\end{table}

\subsection{Optimal Policies and Convergence}\label{ssec:convergence}

\begin{figure*}[h!]
        \centering
        \begin{subfigure}{.5\textwidth}
          \centering
              \includegraphics[width=\textwidth]{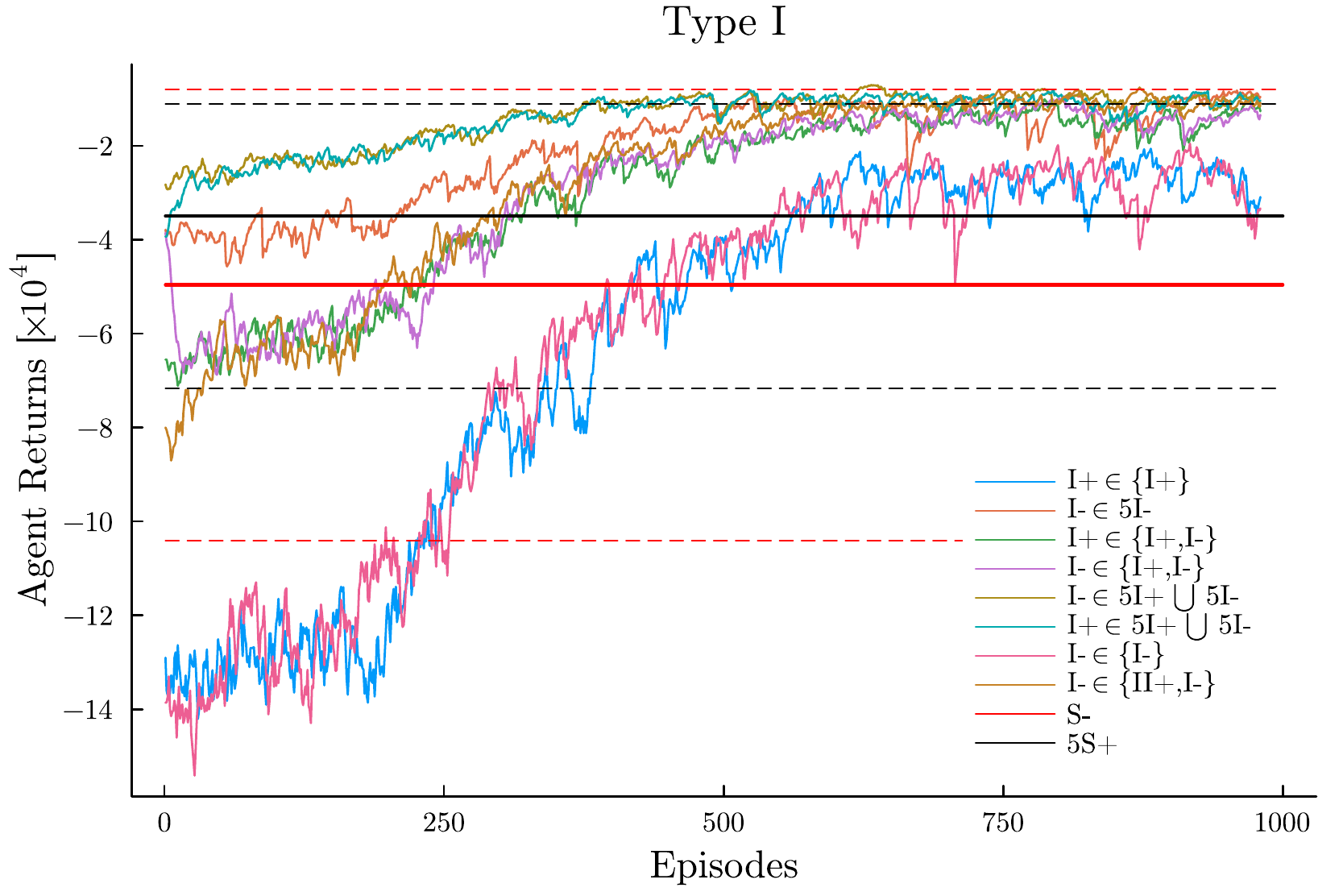}
              \caption{Type I agent return convergence}
              \label{fig:a:return-convergence}
        \end{subfigure}%
        \begin{subfigure}{.5\textwidth}
          \centering
          \includegraphics[width=\textwidth]{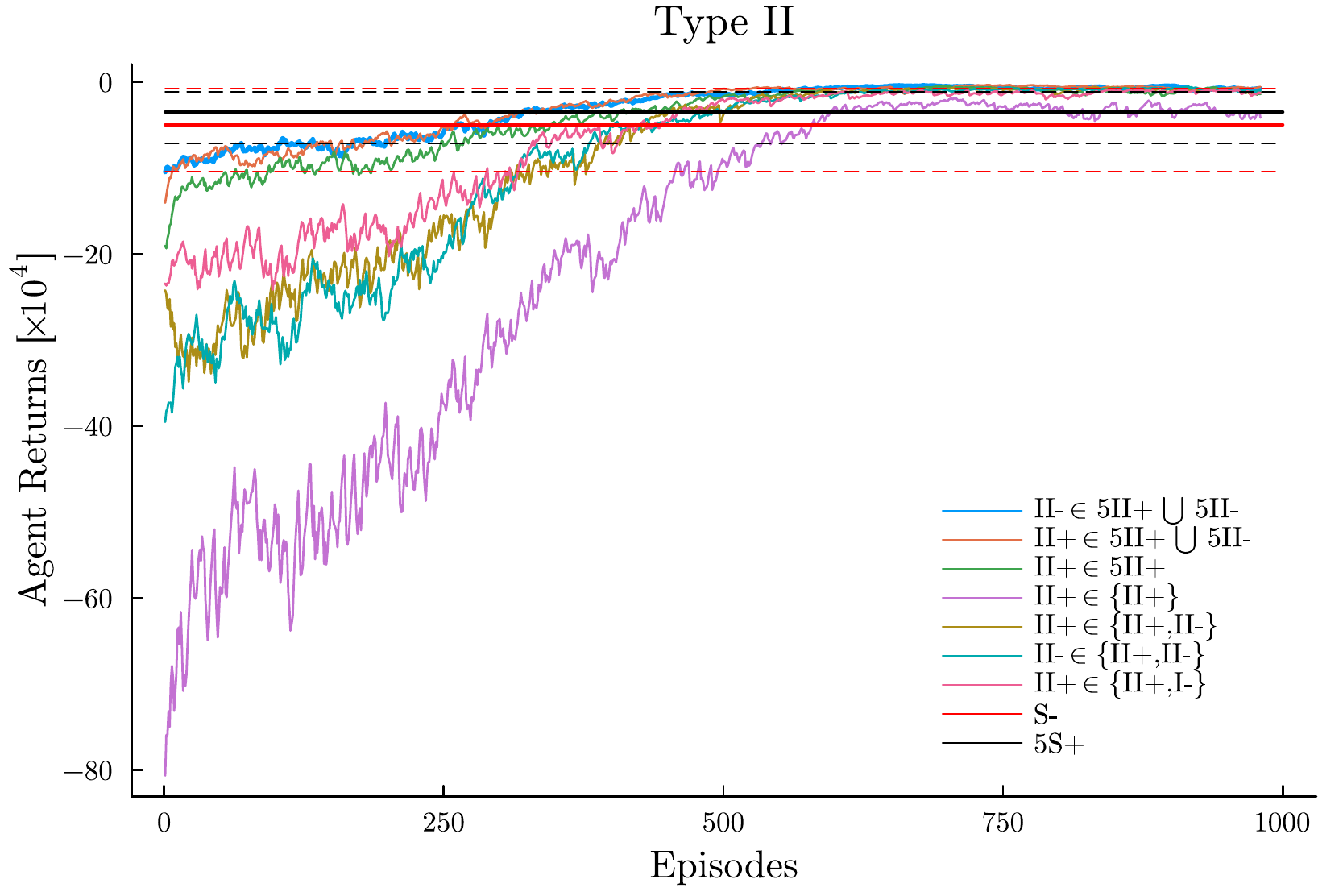}
          \caption{Type II agent return convergence}
          \label{fig:b:return-convergence}
        \end{subfigure}
        \caption{The agent returns are given as a function of the training episodes. This demonstrates how the different agent's rewards converge under training. In Figure \ref{fig:a:return-convergence}, the return rewards for type I agent, both buyers (+) and sellers (-) are shown. Type I agents only trade using market orders.\ref{fig:b:return-convergence} has the same plot, but for agents of type II\,---\,agents that use both market orders and limit orders.}
        \label{fig:return-convergence}
\end{figure*}

To visualise the convergence of the training of learning agents we plot the agent policy returns as a function of the training episodes in Figure  \ref{fig:return-convergence}. Here we have used $1000$ training episodes. Figure \ref{fig:b:return-convergence} shows the return rewards for agent type I, where both buyers (+) and sellers (-) are shown.  Agents are taken from different model configuration sets {\it e.g.} the blue line is a lone buying agent ($I^+$) using only MOs and has similar dynamics to the lone selling agent ($I^-$) given in pink. This shows that the rewards appear to converge under training. Similarly, Figure \ref{fig:b:return-convergence} has the same plot, but for agents of type II\,---\,agents that use both MOs and LOs. Again, we note reasonable evidence of convergence behaviour in the reward function over the 1000 training episodes for the ten different learning agent configurations. The benchmark TWAP agents ($S^\pm$) are included for comparison.

\begin{figure*}[h!]
\begin{subfigure}{.5\textwidth}
    \centering
    \includegraphics[width=\textwidth]{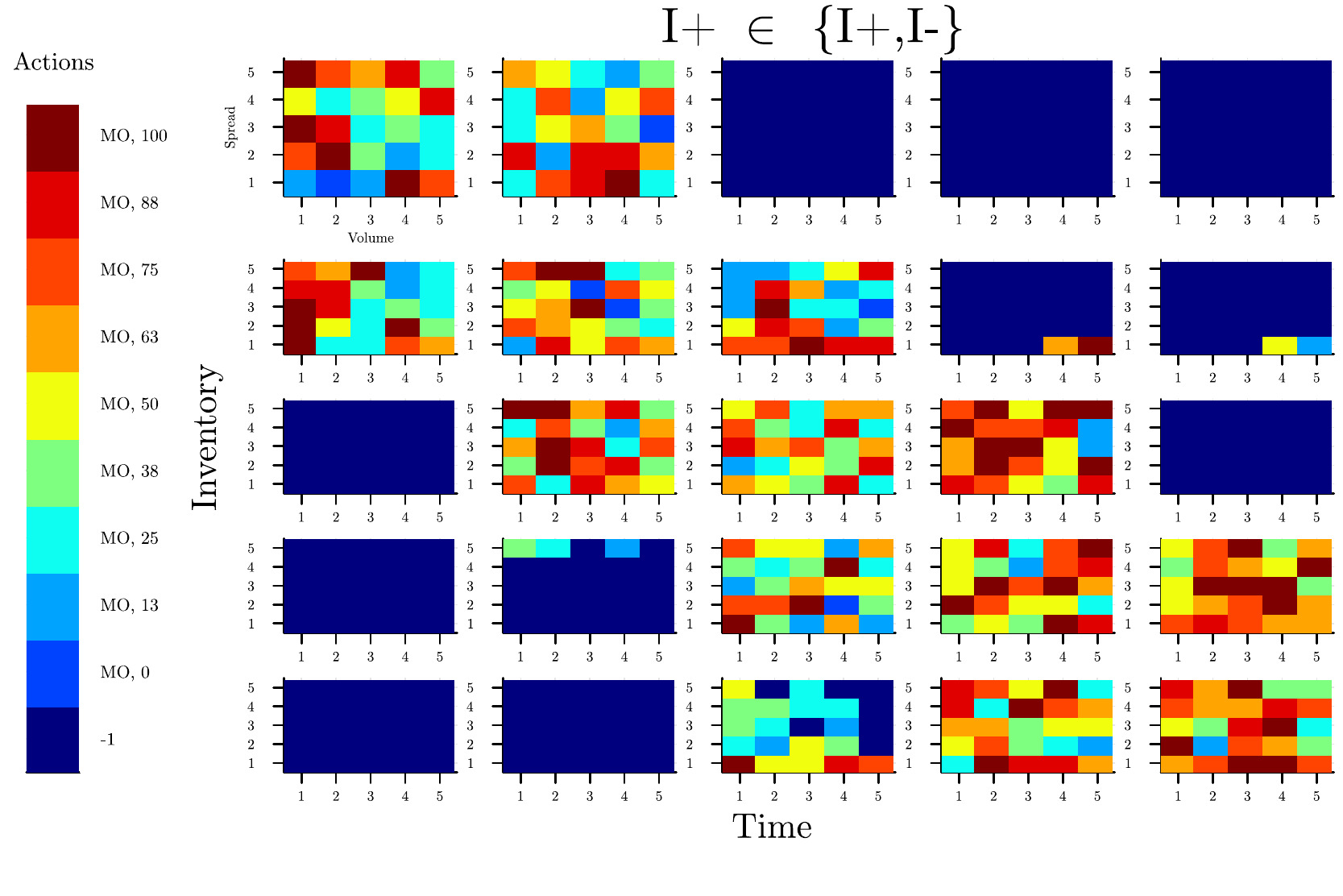}
    \caption{Case 5 buying: Final greedy policy}
    \label{fig:case3-policy-plot-A}
 \end{subfigure}
 \begin{subfigure}{.5\textwidth}
    \centering
     \includegraphics[width=\textwidth]{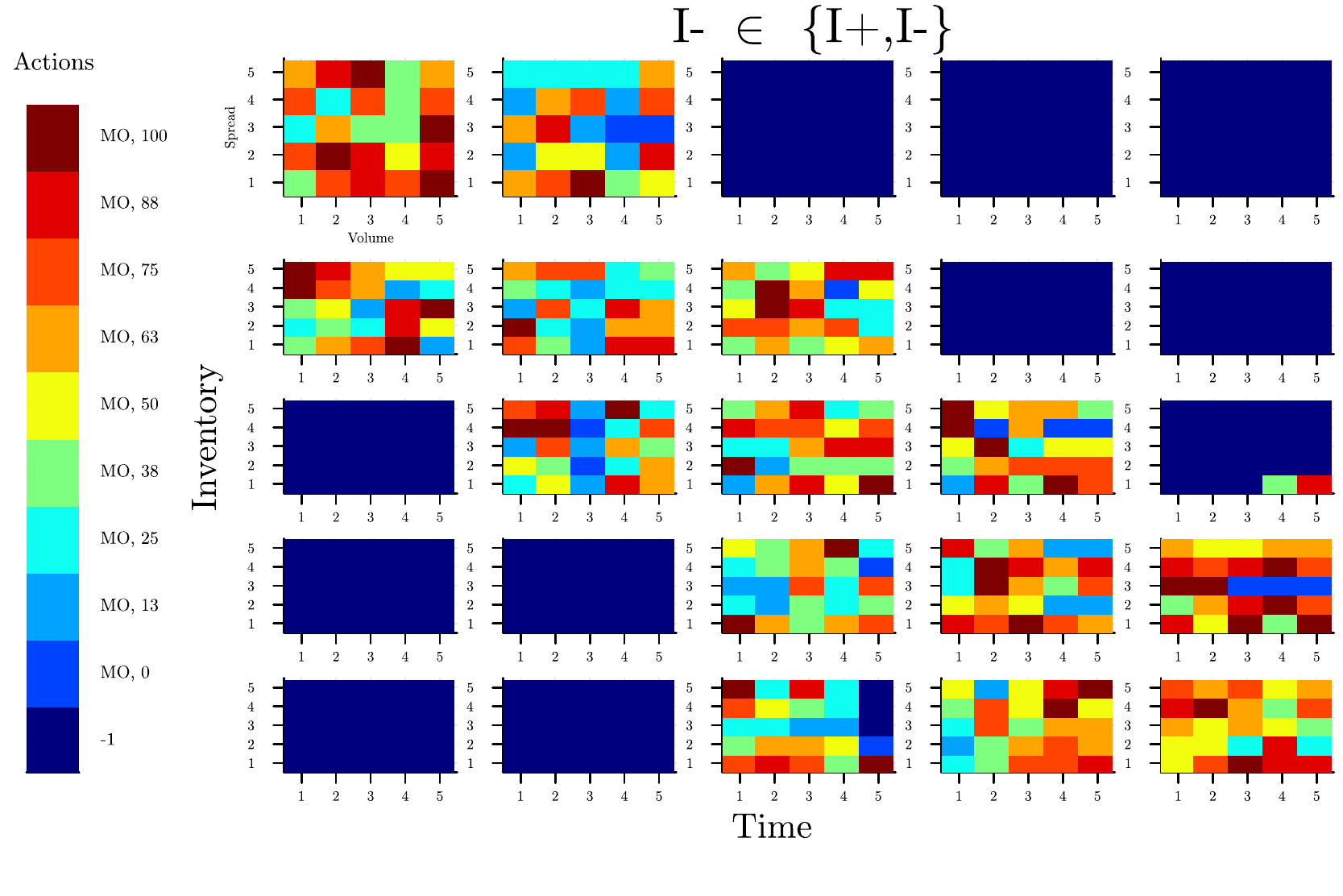}
     \caption{Case 5 selling: Final greedy policy}
     \label{fig:case3-policy-plot-B}
 \end{subfigure}
    \caption{The final $\epsilon$-greedy policy is shown for a type I agent example as a heat map. Here for Case 3 (see Table \ref{tab:RLagentscombinations}) with the buying agent in Figure \ref{fig:case3-policy-plot-A} and the selling agent in \ref{fig:case3-policy-plot-B}. }
    \label{fig:case3-policy-plot}
\end{figure*}

\begin{figure}[h!]
    \centering
    \includegraphics[width=0.5\textwidth]{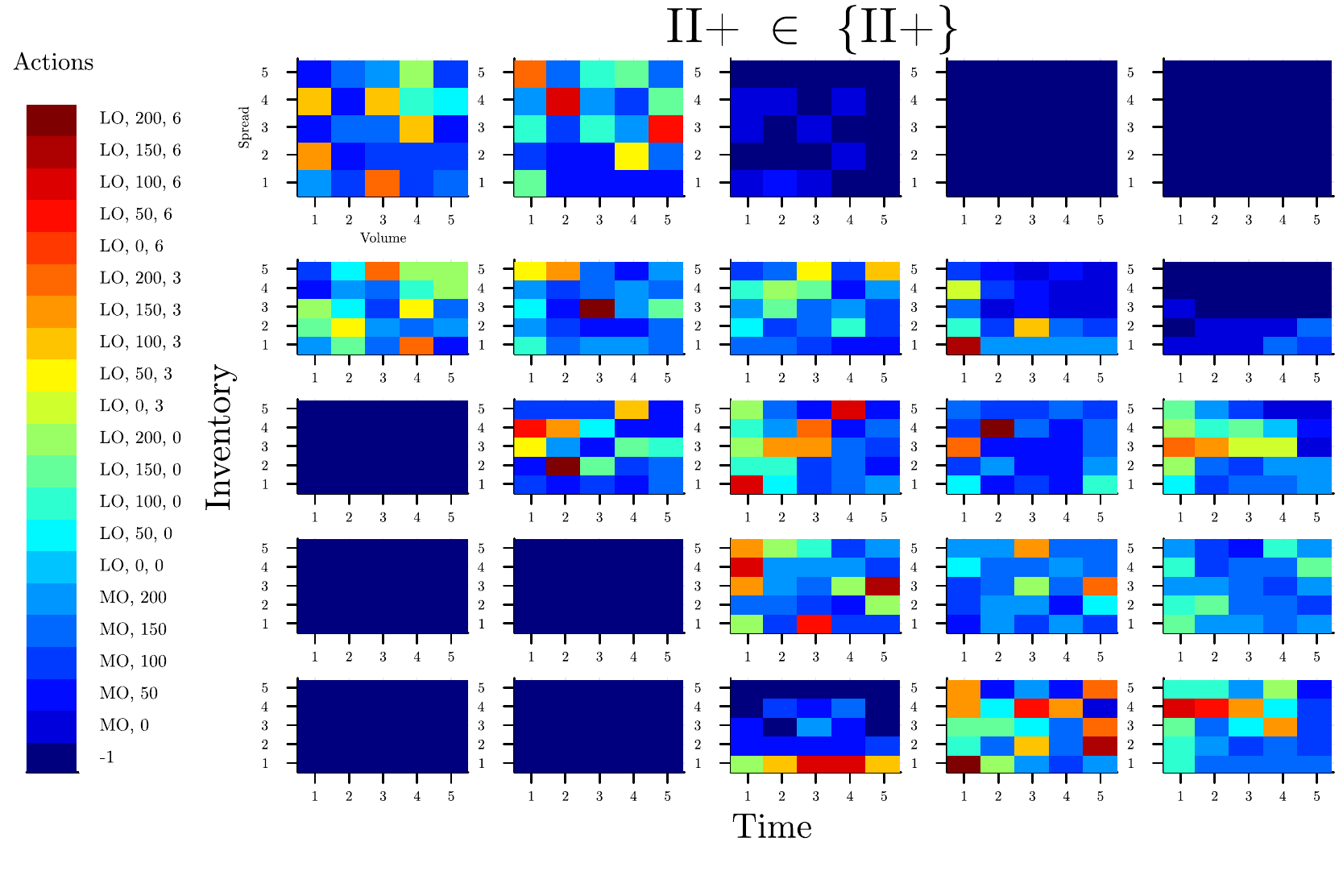}
    \caption{The final greedy policy example for a type II agent as shown for Case 4 (see Table \ref{tab:RLagentscombinations})  as a heat map.}
    \label{fig:case4-policy-plot}
\end{figure}

In Figure \ref{fig:case3-policy-plot} we show the final buying and selling learning agents greedy policy after 1000 training episodes in the training environment \cite{Dicksetal2024ABM} for Case 5 (the case which includes both type $I^+$ and $I^-$ agents together in the environment). The five discrete inventory states increase from bottom to top, and the temporal states increase from left to right across the episode. Within each inventory and time state there are five spread and volume states, represented as heat maps. The specific actions are labelled as colour legend on the left. Action ``-1" represents the state has not been reached, and increasing aggression, via the size label, moving from the bottom to top. Here only MOs are used. The learning agent with a mixture of MOs and LOs is given in the action space example for Case 6 in Figure \ref{fig:case4-policy-plot}. 

In Figure \ref{fig:case4-policy-plot} we have the final greedy policy for Case 6 (type II$^-$) learnt over 1000 training episodes in the ABM training environment. The five inventory states increase from  bottom to top, and the five temporal states from left to right, from the first fifth of the training episode to the last fifth. Within each inventory and time combination a $5 \times 5$ heat map for the spread and volume state combinations is plot. The legend on the left provides the mapping from the number of actions taken for a particular combination to the action and represents the greedy actions taken in a particular state. The Case 4 agent is a combination of MOs and LOs actions with varying aggression.

\section{Exploratory Data Analysis}\label{sec:results}

The measured trade-and-quote (TAQ) data is a single 8-hour day of trading, on 08-07-2019, for a single Johannesburg Stock Exchange (JSE) dual-listed equity, Naspers (NPN.J) \cite{Jericevich2020data}. The data excludes all auctions. The simulated data similarly represents a single period comparable to 8 hour of trading on the basis of the Average Daily Volume (ADV) where the ABM component was calibrated to the estimated moments. The method of simulated moments was used to calibrate the model for the base case (Case 0) and is described in prior work \cite{arxiv2021abm,Dicksetal2024ABM}. 

\subsection{Comparing Stylised Facts}\label{ssec:stylisedfacts}

Table \ref{tab:estimated-moments-table} compares the moments for key configurations. The sample paths used to capture the model variations are comparable to those found in the calibration. The cases are ranked, left to right, on the micro-price fluctuation volatilities and the  Geweke and Porter-Hudak (GPH) statistic \cite{geweke1983estimation} from Table \ref{tab:estimated-moments-table} \,---\ this is a measure of the long-range dependency. We find that all the models have more extreme events than found in the real-world data and that none of the classes suggests evidence of a unit-root. 

Multiple balanced learning agents (Case 5) increase the long-range dependencies (GPH) and mean-reversion (as measured by the Hurst exponent \cite{Hurst1951}), possibly because of interactions. Conversely, Type II agents decrease long-range dependency whilst reducing volatility. We believe this occurs since allowing execution agents to post LOs introduces additional heterogeneity in the order flow and liquidity provision processes, reducing long range dependence, whilst the additional liquidity provided by these agents reduces the occurrence of liquidity shocks, decreasing volatility. From the Hill estimators \cite{nuyts2010inference}, we see that Case 5 has less extreme events (a slower decay in the tail distribution exponent) than Case 1, which may suggest that balanced execution agents ({\it i.e.} an equal number of buying and selling agents) more generally, reduces the tail effects. 


\begin{table*}[h!]
\centering
\begin{tabular}{|l|rl rl rl rl|rl|}
\hline
& \multicolumn{8}{c|}{Simulated} 
& \multicolumn{2}{c|}{Estimated} 
\\
& \multicolumn{2}{c}{Case 5:{\small$\mathrm{ABM}+\mathrm{I}^{\pm}$}}
& \multicolumn{2}{c}{Case 1:{\small$\mathrm{ABM}+\mathrm{S}^-$}}
& \multicolumn{2}{c}{Case 0:{\small ABM}}
& \multicolumn{2}{c|}{Case 6:{\small$\mathrm{ABM}+\mathrm{II}^{+}$}}
& \multicolumn{2}{c|}{JSE:{\small NPN.J}} 
\\
Moment 
& $\boldsymbol{m}^s$ 
& \small[97.5\% CI] 
& $\boldsymbol{m}^s$ 
& \small[97.5\% CI] 
& $\boldsymbol{m}^s$ 
& \small[97.5\% CI] 
& $\boldsymbol{m}^s$ 
& \small[97.5\% CI] 
& $\boldsymbol{m^e}$ 
& \small[97.5\% CI] \\
\hline
Mean & 0 & - & 0 & -  & 0 & - & 0 & - & 0 & - 
\\
Std$\times 10^{-4}$ 
& 4.60 & {\small[3.90,5.30]} 
& 4.04 & {\small[3.36,4.75]} 
& 2.31 & {\small[1.61,3.02]} 
& 1.48 & {\small[0.78,2.18]} 
& 1.39 & {\small[1.19,1.59]} 
\\
KS                      
& 0.22 & {\small[0.16,0.28]}  
& 0.16 & {\small[ 0.16,0.27]} 
& 0.18 & {\small[0.12,0.23]}  
& 0.27 & {\small[0.21,0.32]}  
& 0.00 & {\small[-0.01,0.01]} 
\\
Hurst 
& 0.29 & {\small[0.22,0.35]} 
& 0.33 & {\small[0.26,0.40]} 
& 0.40 & {\small[0.33,0.46]} 
& 0.38 & {\small[0.31,0.44]} 
& 0.47 & {\small[0.41,0.52]} 
\\
GPH 
& 0.69 & {\small[0.57,0.82]} 
& 0.62 & {\small[0.49,0.74]} 
& 0.51 & {\small[0.39,0.63]} 
& 0.42 & {\small[0.29,0.54]} 
& 0.44 & {\small[0.30,0.59]}
\\
ADF 
& -154 & {\small[-158,-151]} 
& -158 & {\small[-161,-154]} 
& -148 & {\small[-152,-145]} 
& -167 & {\small[-171,-164]} 
& -136 & {\small[-140,-133]} 
\\ 
GARCH 
& 1.31 & {\small[1.24,1.38]} 
& 0.95 & {\small[0.87,1.02]} 
& 0.99 & {\small[0.91,1.05]} 
& 1.06 & {\small[0.99,1.13]} 
& 1,00 & {\small[0.96,1.03]} 
\\ 
Hill 
& 0.83 & {\small[0.41,1.25]} 
& 1.39 & {\small[0.97,1.81]} 
& 1.24 & {\small[0.82,1.66]} 
& 1.25 & {\small[0.83,1.67]} 
& 1.99 & {\small[1.72,2.26]} 
\\ 
\hline
\end{tabular}
\caption{Simulated moments (using the calibrated model for the environment) and estimated moments.  These are the same used in prior work \cite{Dicksetal2024ABM}. The estimated moments are from the market data \cite{Dicksetal2024ABM}. The selected simulated moment set is ordered on decreasing micro-price volatility (Std.) left to right. The mean and standard deviation (Std) are the usual sample moments, the remaining are as follows, Kolmogorov-Smirnov (KS) statistic \cite{Massey1951}, Hurst exponent \cite{Hurst1951}, Geweke and Porter-Hudak (GPH) statistic \cite{geweke1983estimation}, Augmented Dickey-Fuller (ADF) statistic \cite{DickeyFuller1979}, the sum of the parameters of a GARCH(1,1) model \cite{winker2007objective}, and the improved Hill estimator \cite{nuyts2010inference}. The cases left to right from Table \ref{tab:RLagentscombinations}: Case 5 is the ABM environment with both a buying and selling agent trading in MO's, Case 1 has the ABM combined with the benchmark TWAP trader, Case 0 is the ABM environment alone, and Case 6, that ABM combined with a single buying agent trading in both LO's and MO's.}
\label{tab:estimated-moments-table}
\end{table*}

\begin{figure*}[htbp]
    \centering
    \begin{tikzpicture}
        \begin{axis}[
            name=plot1,
            title={Buyer-Initiated Trades},
            xlabel={$\omega$},
            ylabel={$\delta p^{*}$},
            tick label style={font=\tiny},
            grid=none,
            width=0.45\textwidth,
            xmode=log,
            ymode=log,
            axis lines=left,
            enlargelimits=0.1
            ]        
            \pgfplotsset{cycle list={%
                {blue, densely dashed},
                {blue, dashed},
                {blue, dotted},
                {blue, dashdotted},
                {blue, densely dotted},
                {red, densely dashed},
                {red, dashed},
                {red, dotted},
                {red, dashdotted}
            }}
            \foreach \i in {3, 5, 6, 7, 8, 9, 10, 11, 12} {
                \addplot+[line width=1pt] table[x=omega_\i, y=delta_price_\i, col sep=comma] {Figures/Stylised_Facts/buyer_pi_curves.csv};
            }
        \end{axis}

        \begin{axis}[
            at={(plot1.south east)}, 
            xshift=2cm, 
            title={Seller-Initiated Trades},
            xlabel={$\omega$},
            ylabel={$\delta p^{*}$},
            tick label style={font=\tiny},
            grid=none,
            width=0.45\textwidth,
            xmode=log,
            ymode=log,
            axis lines=left,
            enlargelimits=0.1,
            legend style={at={(-0.15,-0.15)}, anchor=north, draw=none, legend columns =9}
            ]        
            \pgfplotsset{cycle list={%
                {blue, densely dashed},
                {blue, dashed},
                {blue, dotted},
                {blue, dashdotted},
                {blue, densely dotted},
                {red, densely dashed},
                {red, dashed},
                {red, dotted},
                {red, dashdotted}
            }}
            \foreach \i in {3, 5, 6, 7, 8, 9, 10, 11, 12} {
                \addplot+[line width=1pt] table[x=omega_\i, y=delta_price_\i, col sep=comma] {Figures/Stylised_Facts/seller_pi_curves.csv};
            }
            \addlegendentry{$I^{-}$} 
            \addlegendentry{$5I^{-}$} 
            \addlegendentry{$I^{+}$} 
            \addlegendentry{$I^{+}, I^{-}$} 
            \addlegendentry{$5I^{+}, 5I^{-}$} 
            \addlegendentry{$II^{+}$} 
            \addlegendentry{$II^{-}, II^{+}$} 
            \addlegendentry{$5II^{+}$} 
            \addlegendentry{$5II^{+}, 5II^{-}$} 
        \end{axis}
    \end{tikzpicture}
    \caption{Price impact plots for the buyer (left) and seller (right) initiated trades for different cases of learning agents. Type I (blue) agents use only market orders to execute parent orders, whilst Type II (red) agents use market and limit orders. The lower price impact of Type II agents suggests that limit orders can be used to take advantage of opportunities created by market flow and changes in the spread.}
    \label{fig:price-impact}
\end{figure*}
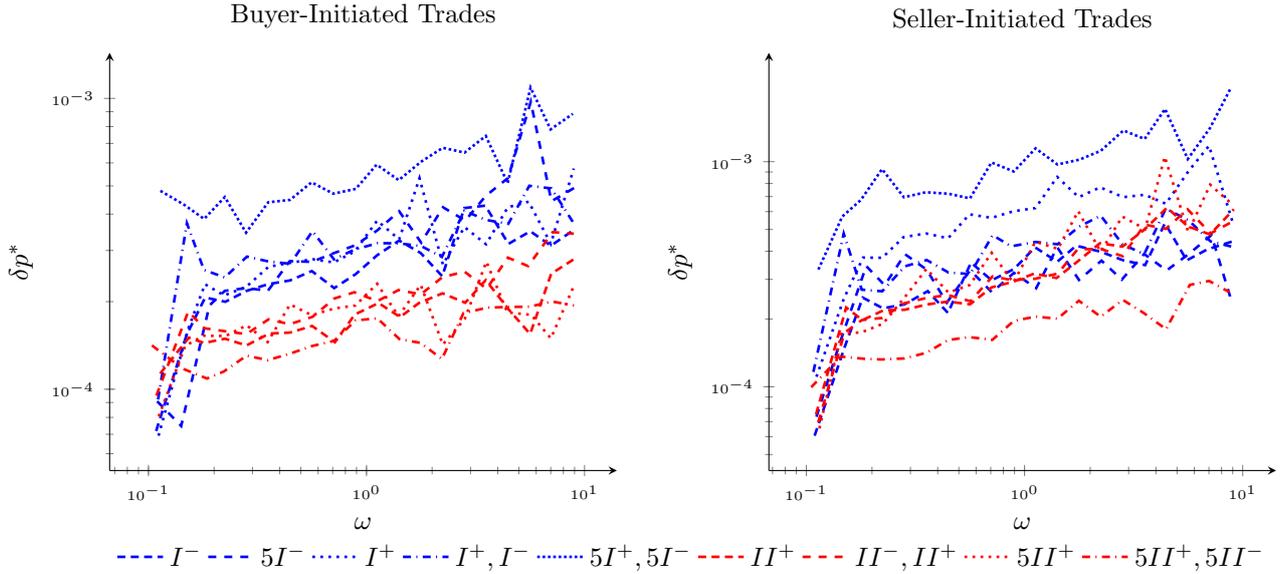

\begin{figure}[h!]
    \centering
    \begin{tikzpicture}
        \begin{axis}[
            title={Type I and II Absolute Log-Return ACF},
            grid=none,
            axis lines=left,
            enlargelimits=0.1,
            width=0.5\textwidth,
            legend style={font=\small, at={(0.95,0.95)}, anchor=north east, legend columns=2, draw=none, reverse legend},
            cycle list={%
                {blue, densely dashed},
                {blue, dashed},
                {blue, dotted},
                {blue, dashdotted},
                {blue, densely dotted},
                {red, densely dashed},
                {red, dashed},
                {red, dotted},
                {red, dashdotted}
            }
            ]
            \foreach \i in {3, 5, 6, 7, 8, 9, 10, 11, 12} {
                \addplot+[line width=1pt] table[x=index, y=\i, col sep=comma] {Figures/Stylised_Facts/logret_acfs.csv};
            }
            \addlegendentry{$I^{-}$} 
            \addlegendentry{$5I^{-}$} 
            \addlegendentry{$I^{+}$} 
            \addlegendentry{$I^{+}, I^{-}$} 
            \addlegendentry{$5I^{+}, 5I^{-}$} 
            \addlegendentry{$II^{+}$} 
            \addlegendentry{$II^{-}, II^{+}$} 
            \addlegendentry{$5II^{+}$} 
            \addlegendentry{$5II^{+}, 5II^{-}$} 
            
        \end{axis}
    \end{tikzpicture}
    \caption{Auto-Correlation Function (ACF) plots of the absolute value of the mid-price returns comparing Type I (blue) and Type II (red) cases. Type I agents have non-trivial auto-correlations, while Type II agents suppress autocorrelations. This is indicative that the more complex Type II agents reduce regularity in potentially both the order flow and the liquidity provision processes.}
    \label{fig:absreturn-autocorr}
\end{figure}
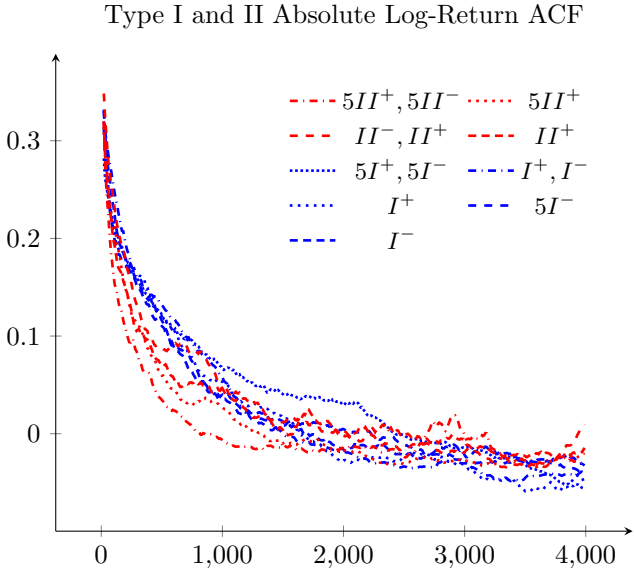

\subsection{Persistence of Orderflow}\label{ssec:acftradesigns}

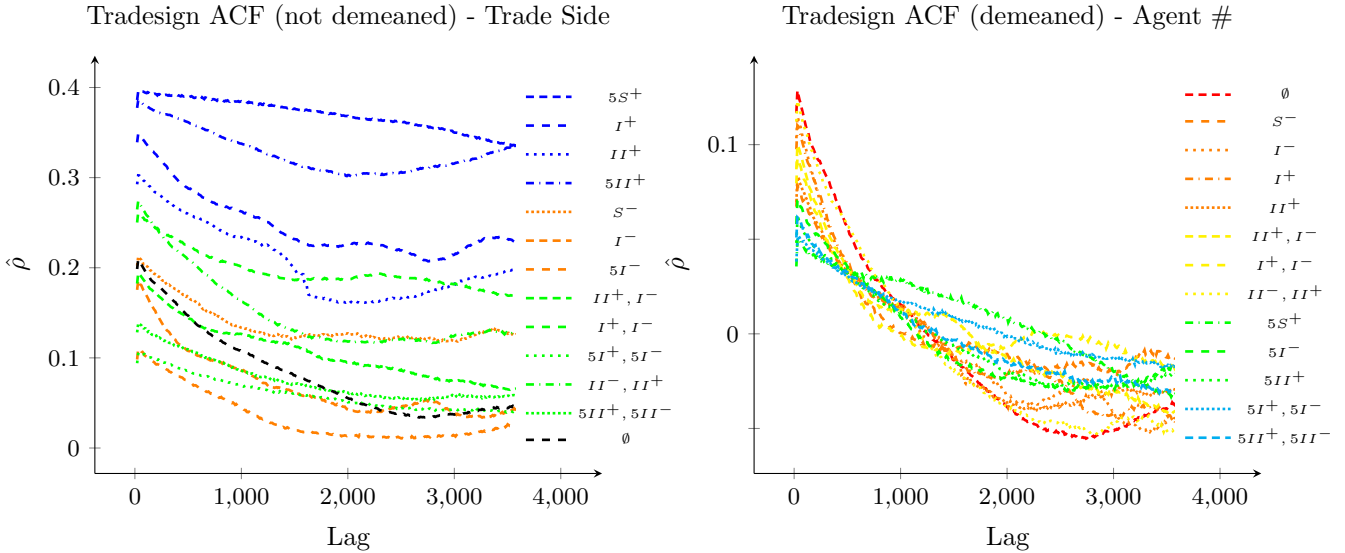
\begin{figure*}[h!]
    \centering
    \begin{tikzpicture}
        \begin{axis}[
            name=plot1,
            title={Tradesign ACF (not demeaned) - Trade Side},
            xlabel={Lag},
            ylabel={$\hat{\rho}$},
            tick label style={font=\small},
            grid=none,
            width=0.45\textwidth,
            axis lines=left,
            enlargelimits=0.1,
            legend style={font=\tiny, at={(1,0.95)}, anchor=north, legend columns=1, draw=none}
            ]          
            \pgfplotsset{cycle list={%
                {blue, densely dashed},
                {blue, dashed},
                {blue, dotted},
                {blue, dashdotted},
                {orange, densely dotted},
                {orange, densely dashed},
                {orange, dashed},
                {green, densely dashed},
                {green, dashed},
                {green, dotted},
                {green, dashdotted},
                {green, densely dotted},
                {black, dashed}
            }}
            \foreach \i in {2, 6, 9, 11, 1, 3, 5, 4, 7, 8, 10, 12, 0} {
                \addplot+[line width=1pt] table[x=index, y=\i, col sep=comma] {Figures/Stylised_Facts/acfs_side.csv};
            }
            \addlegendentry{$5S^{+}$} 
            \addlegendentry{$I^{+}$} 
            \addlegendentry{$II^{+}$} 
            \addlegendentry{$5II^{+}$} 
            \addlegendentry{$S^{-}$} 
            \addlegendentry{$I^{-}$} 
            \addlegendentry{$5I^{-}$} 
            \addlegendentry{$II^{+}, I^{-}$} 
            \addlegendentry{$I^{+}, I^{-}$} 
            \addlegendentry{$5I^{+}, 5I^{-}$} 
            \addlegendentry{$II^{-}, II^{+}$} 
            \addlegendentry{$5II^{+}, 5II^{-}$} 
            \addlegendentry{$\emptyset$} 
        \end{axis}

        \begin{axis}[
            at={(plot1.south east)}, 
            xshift=2cm, 
            title={Tradesign ACF (demeaned) - Agent \#},
            xlabel={Lag},
            ytick = {-0.05, 0, 0.05, 0.1},
            yticklabels={, 0, , 0.1},
            ylabel={$\hat{\rho}$},
            tick label style={font=\small},
            grid=none,
            width=0.45\textwidth,
            axis lines=left,
            enlargelimits=0.1,
            legend style={font=\tiny, at={(1,0.95)}, anchor=north, legend columns=1, draw=none},
            ]         
            \pgfplotsset{cycle list={%
                {red, densely dashed},
                {orange, dashed},
                {orange, dotted},
                {orange, dashdotted},
                {orange, densely dotted},
                {yellow, densely dashed},
                {yellow, dashed},
                {yellow, dotted},
                {green, dashdotted},
                {green, dashed},
                {green, dotted},
                {cyan, densely dotted},
                {cyan, densely dashed}
            }}
            \foreach \i in {0, 1, 3, 6, 9, 4, 7, 10, 2, 5, 11, 8, 12} {
                \addplot+[line width=1pt] table[x=index, y=\i, col sep=comma] {Figures/Stylised_Facts/acfs_tradesigns.csv};
            }
            \addlegendentry{$\emptyset$} 
            \addlegendentry{$S^{-}$} 
            \addlegendentry{$I^{-}$} 
            \addlegendentry{$I^{+}$} 
            \addlegendentry{$II^{+}$} 
            \addlegendentry{$II^{+}, I^{-}$} 
            \addlegendentry{$I^{+}, I^{-}$} 
            \addlegendentry{$II^{-}, II^{+}$} 
            \addlegendentry{$5S^{+}$} 
            \addlegendentry{$5I^{-}$} 
            \addlegendentry{$5II^{+}$} 
            \addlegendentry{$5I^{+}, 5I^{-}$} 
            \addlegendentry{$5II^{+}, 5II^{-}$} 
            
        \end{axis}
    \end{tikzpicture}
    \caption{Auto-Correlation Functions (ACF) of tradesigns where the tradesigns are not demeaned (left) and demeaned (right), respectively. On the left we observe that the level of the ACF reflects the imbalance in the number of execution agents that are buying vs. selling. To demonstrate this we order the plot colours by the trade signs - buying (+), to selling (-), to increasing mixtures of buying and selling ($\pm$). This shows that the level of the ACF increases with more buying agents (blue), is moderated by including both buying and selling agents (green), and decreases by including selling agents only (orange). The directional changes in the ACF most likely reflect that of the environment ABM (black) which was by design calibrated in an upward-trending market. On the right, we demean the ACF so that the imbalance in the number of agents does not visually dominate the plot. We then order the plot colour by the number of agents. We find that increasing the number of agents (red is lowest, cyan is greatest) decreases the rate of decay of the ACF. }
    \label{fig:acf-tradesigns}
\end{figure*}

The ACF of tradesigns reflects the persistence in the direction of the MO flow\footnote{The ACFs are not on demeaned data because the data is ordinal.}. Due to potentially large sampling variation, analysis of a single path is not indicative of general behaviour. On the other hand, averaging sample ACFs tends to conceal interesting differences in the distribution of sample paths between the different cases. Thus, we consider average sample ACFs and the individual sample paths to explain the ABM behaviour, and discuss broad phenomena observed from their plots. 

Firstly, as seen in Figure \ref{fig:acf-tradesigns} (left), the average level of ACF reflects the number and (net) direction of the execution agents. The ABM was calibrated by design in an upward-trending market, and hence the environment base case produces sample paths that are slightly more biased towards buying (Case 0). Adding execution agents increases the persistence of the trade signs, and hence the ACF, simply by increasing the proportion of buys in the order flow. Conversely, including selling agents in the ABM results in an increase in activity in opposition to the original orderflow, thus decreasing the ACFs. Including both selling and buying agents produces ACFs that tend to lie between the ACFs of the buying and selling agents, respectively. The tradesign ACF plot is ordered by colours on the tradesign to make this more apparent.

However,  as seen in Figure \ref{fig:acf-tradesigns} (right), we believe the observed increases or decreases in demeaned ACF, largely reflect the bias of direction of order flow in the ABM. If the ABM was biased towards selling, we would observe the effects in the opposite direction observed here. Nonetheless, we observe that the change in level from including agents of a given type becomes more pronounced with the number of agents, as these execution agents start to dominate the order flow over the other classes. To show this we ordered the plot colours on the number of execution agents interacting with the environment.

Secondly, as seen in Figure \ref{fig:acf-tradesigns}, following the inclusion of execution agents, the ABM produces, with greater frequency, sample paths where the cumulative direction of the order flow persists for very long periods. The corresponding sample ACFs are linear with very slow decay. This is especially evident in the minimally intelligent cases (TWAPs), where ACFs become increasingly linear with greater frequency as the number of agents increases. Surprisingly, the presence of sample paths with persistent cumulative order flow is also evident (although to a lesser degree) when the number of buying and selling agents is equal. However, the remaining sample paths may be nonlinear, with several changes in the direction of cumulative order flow with corresponding nonlinear ACFs, which can produce nonlinear behaviour in the average ACFs. 

Furthermore, when comparing a single selling TWAP agent (Case 1) with, for example, five buying TWAP agents (Case 2) competing for liquidity, it is noted that as the number of optimal execution agents increases from one to five, the cumulative order flow becomes increasingly uniform, leading to slowly decaying monotonic ACFs. 

We observe that increasing the number of agents tends to decrease the rate of change of the ACF and is most evident at low lags (less than 1000 events). In the base case, the orderflow arises from the interaction of fundamentalists and chartists, which tends to have higher autocorrelations at shorter horizons, most likely due to minority game dynamics. In contrast, the order flow increasingly reflects the activity of execution agents as their numbers increase, which tends to result in linear slow decaying ACFs, and decrease the decay rate of ACF. 

\subsection{Price Impact}\label{ssec:priceimpact}

The buyer and seller initiated price impact curves are shown in Figure \ref{fig:price-impact}. Price impact is defined as the instantaneous change to the mid-price following a trade, which depends on the shape of the orderbook at the time of trade. The greater the amount of liquidity, particularly at prices at and close to best quotes, the lower the price impact. Conversely, reducing the available liquidity will increase price impact. 

Liquidity supply is dynamic and is reduced by liquidity-taking agents (and order cancellations), and is increased by liquidity-providing agents. Thus, differences between cases can be attributed to the differences in the trading and liquidity provision behaviour of the different agent classes. As with the analysis of tradesigns, the trading and liquidity provision behaviour is path-dependent and hence has a large degree of sampling variation, which we try to eliminate by analysing average price impact curves. 

The clearest pattern that we observe in Figure \ref{fig:price-impact} is that cases with Type II agents have lower price impact than Type I agents. Type II agents can use LOs in place of MOs to execute their parent orders, thereby reducing liquidity-taking behaviour while increasing liquidity provision, both of which result in greater average available liquidity. The difference in the relative level of the buyer and seller initiated price impact can be explained to be due to the overall trend in the market \---\ which is upward.


\subsection{Memory in Absolute Returns}\label{ssec:acfabsreturn}

Figure \ref{fig:absreturn-autocorr} shows the ACF of absolute returns. This measures persistence in the size of returns calculated from micro-prices \footnote{The ACF of the absolute returns uses demeaned data.}, reflecting the dynamics of the top of the order book. By definition, a change in the micro-price arises from a change in the top-of-book price and/or top-of-book volume. Changes in these quantities are due to events: a trade, a new limit order or a cancellation, and how these quantities change depends on the shape of the order book. The different event-types are mutually exclusive and have their own processes defined by volume, price, and relative frequency. 

Thus, the ACF of absolute returns encapsulates the behaviour of a wider array of market variables, in comparison to the ACF of tradesigns and price impact curves, making any observed patterns remarkable and worthy of attention, but difficult to interpret.

In Figure \ref{fig:absreturn-autocorr} we observe two patterns which appear to support the hypothesis that the variation in liquidity demand in excess of liquidity determines the decay in the ACF of absolute returns. Firstly, the ACF decays at the slowest rate when five buying TWAP agents (Case 2) are added to the base ABM, since this reduces variation in the volume and frequency of MOs. Furthermore, we see more generally that the ACFs of Type II agents decay faster than Type I agents, because the ability to post LOs of varying size and depth introduces further variation in order flow and liquidity processes, which is temporally uncorrelated due to changes in the market's state. However, limitations in the data prevented convincing support or falsification of this hypothesis.

\subsection{The missing complexity}\label{ssec:complexity}

Although we are able to recover many of the stylised facts (See Table \ref{tab:estimated-moments-table} and figures \ref{fig:price-impact}, \ref{fig:absreturn-autocorr} and \ref{fig:acf-tradesigns}) we are not able to fully recover sufficient model complexity relative to the measure real-world data. This is shown in Figure \ref{fig:complexitydifferences} where the empirical data from the JSE test data is given with confidence intervals (red). The training environment is shown in black with confidence intervals. The learning agent configurations use the type labelling from Table \ref{tab:RLagentscombinations}. 

We notice that the none of the model configurations are able to capture the full complexity of the real-world data, and all have dimensions less than at least $2$. The single agents tend to have dimensions slightly greater than that of the ABM describing the learning environment, and the many agent configurations are slightly less. This is elaborated in Table \ref{tab:dimensiondifferences}. Table \ref{tab:dimensiondifferences} gives the relative differences of the model configurations. The agent case (first column), RL agents types (second column) and $\Delta D$, the difference between fractal dimensions of the different configurations, and the ABM's fractal dimensions, as averaged across the embedding dimension (see Figure \ref{fig:embeddingdimensions}) are sorted on the number of RL agents interacting with the ABM.

\begin{figure*}[ht!]
        \centering
        \begin{subfigure}{.3\textwidth}
          \centering
            \includegraphics[width=\textwidth]{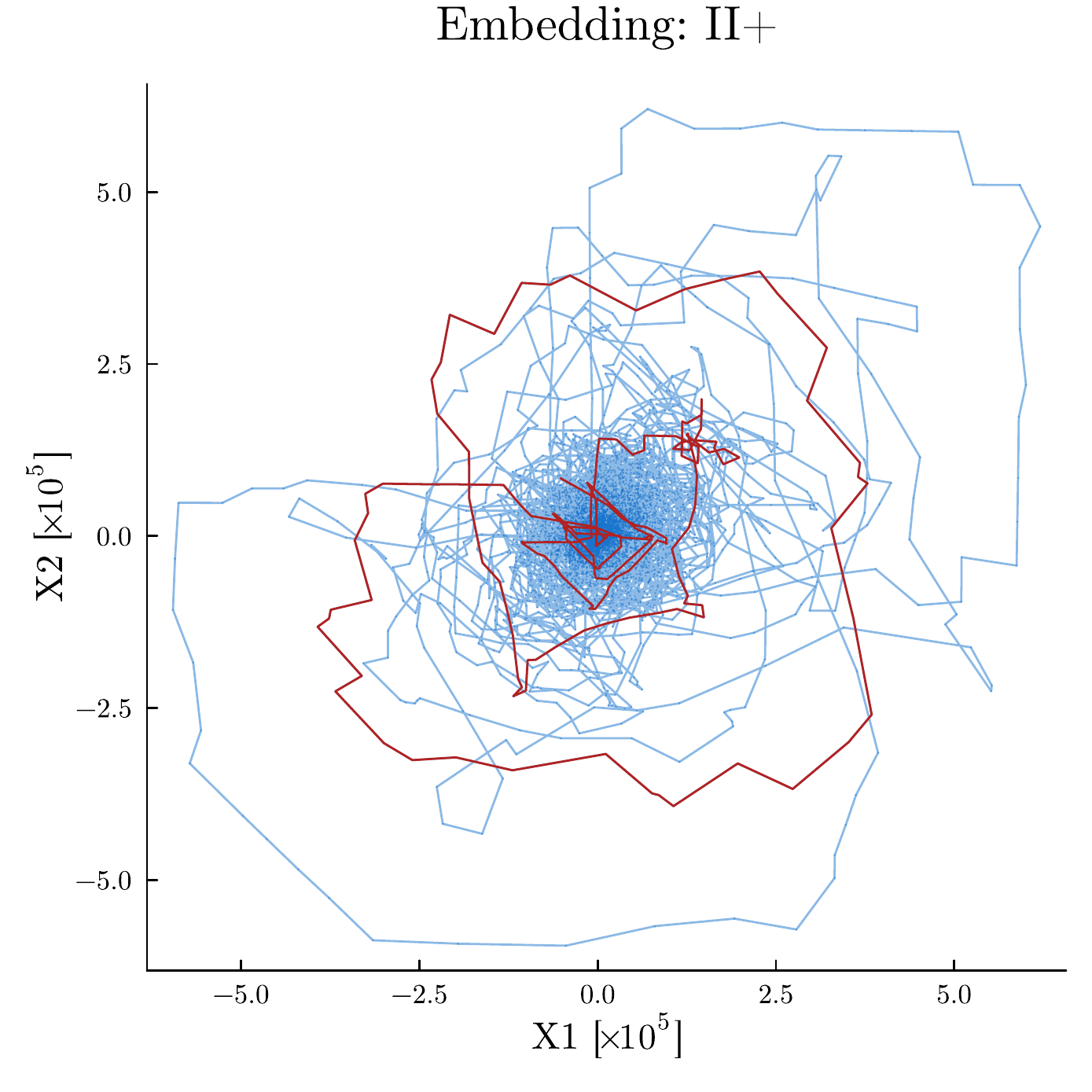}
              \caption{Case 6: phase space plot}
              \label{fig:phase-space-plots-a}
        \end{subfigure}%
        \begin{subfigure}{.3\textwidth}
          \centering
          \includegraphics[width=\textwidth]{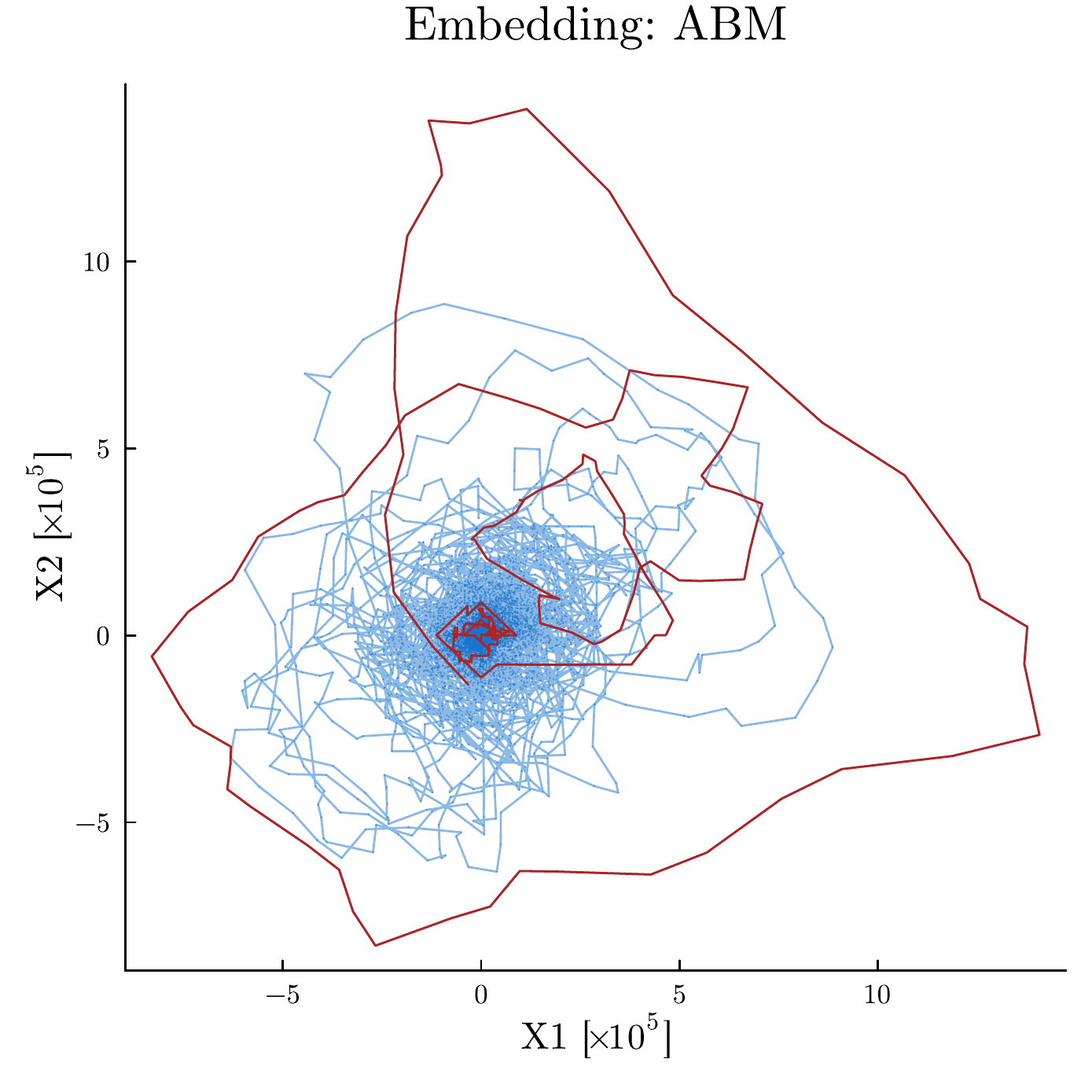}
          \caption{Case 0: ABM phase space}
          \label{fig:phase-space-plots-c}
        \end{subfigure}
                \begin{subfigure}{.3\textwidth}
          \centering
          \includegraphics[width=\textwidth]{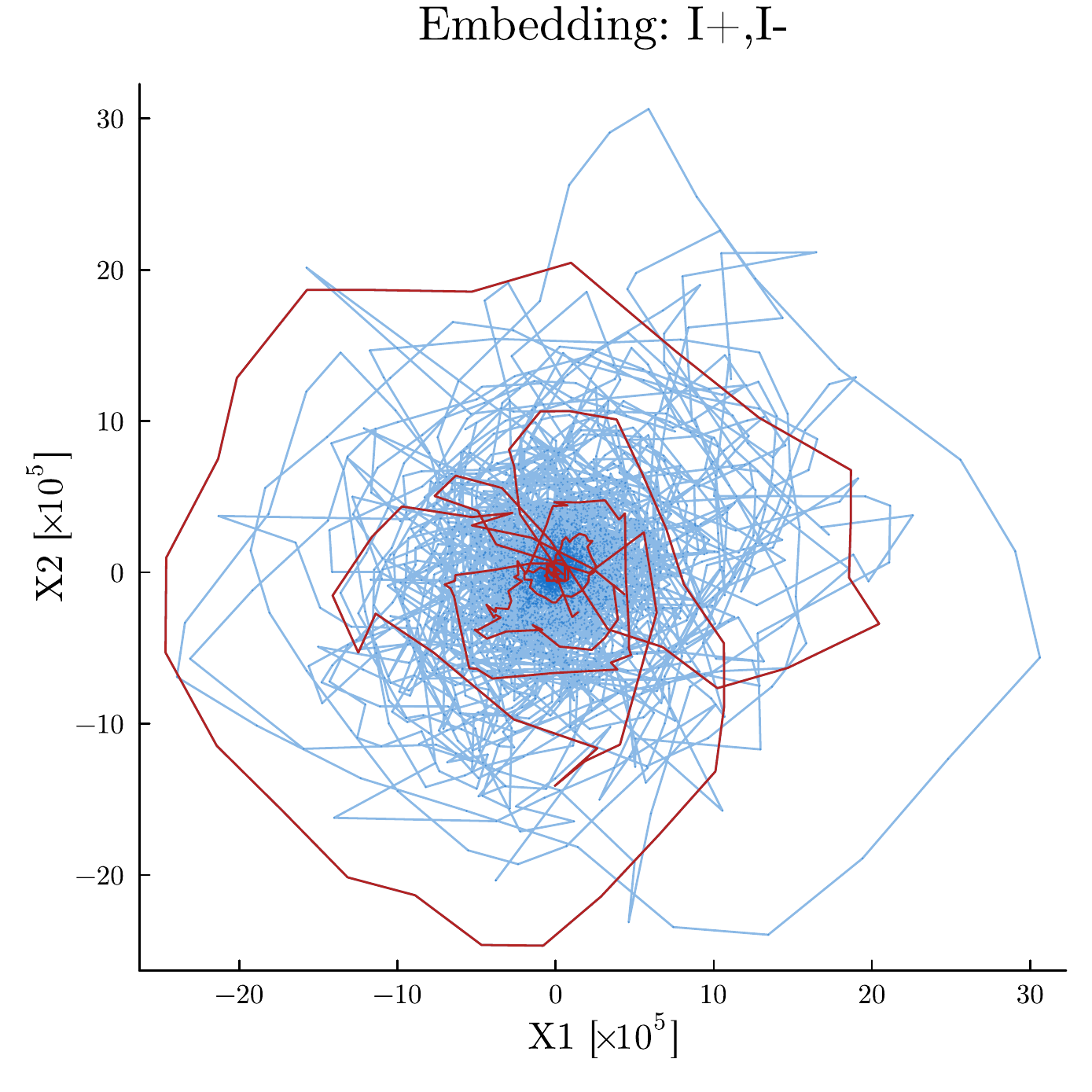}
          \caption{Case 5: phase space plot}
          \label{fig:phase-space-plots-b}
        \end{subfigure}%
        
         \begin{subfigure}{.3\textwidth}
          \centering
            \includegraphics[width=\textwidth]{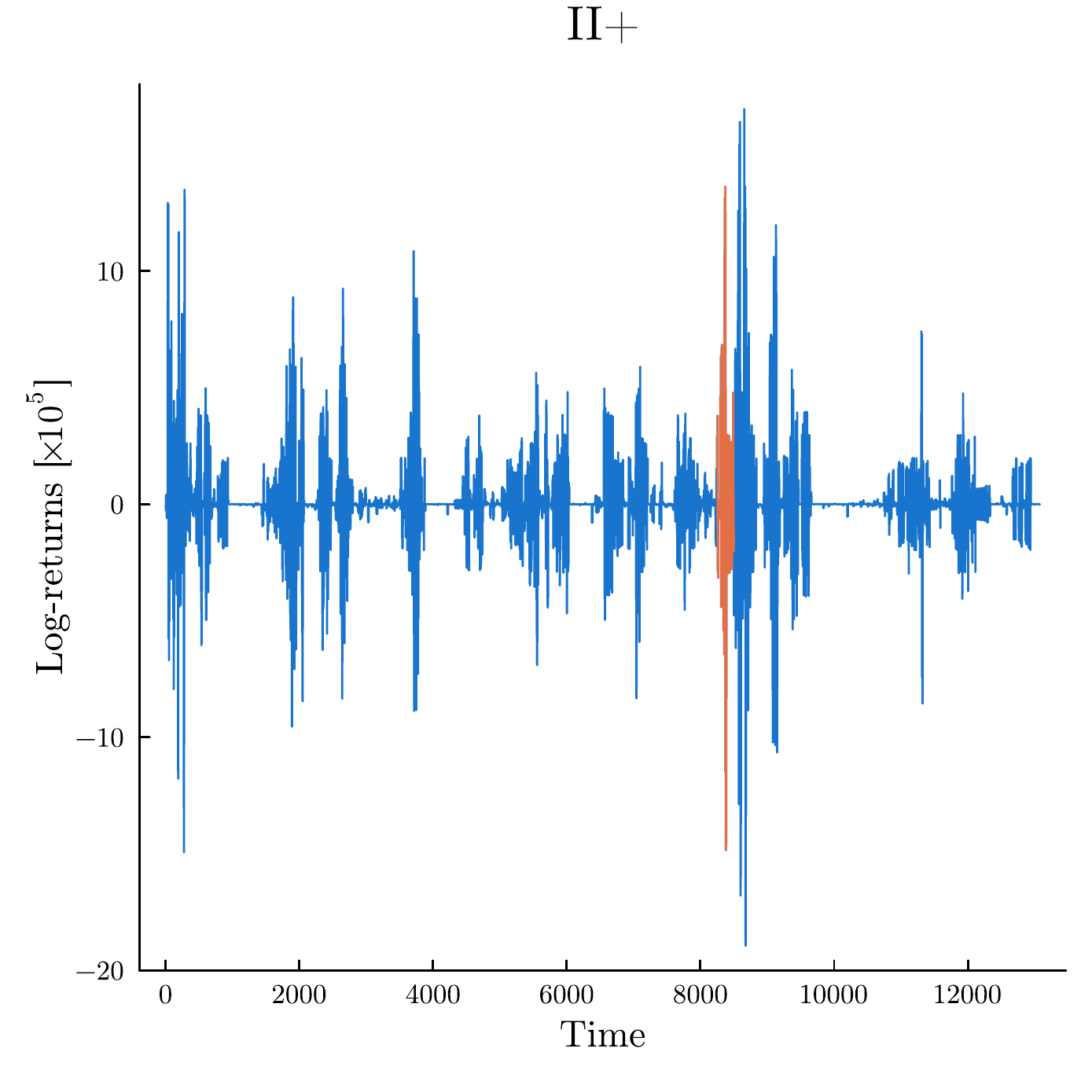}
              \caption{Case 6: micro-price fluctuations}
              \label{fig:time-series-plots-d}
        \end{subfigure}%
        \begin{subfigure}{.3\textwidth}
          \centering
          \includegraphics[width=\textwidth]{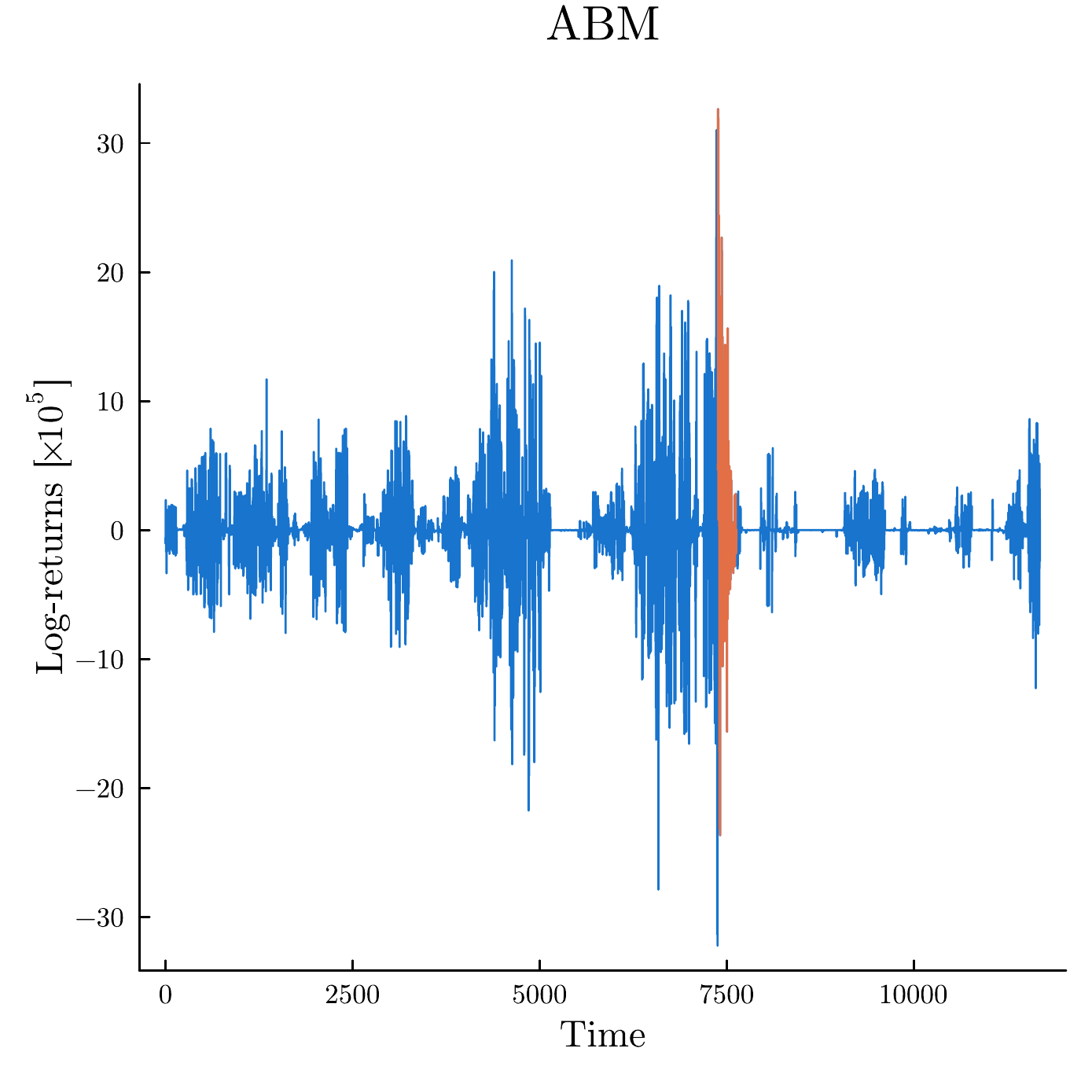}
          \caption{Case 0: micro-price fluctuations}
          \label{fig:time-series-plots-f}
        \end{subfigure}
                \begin{subfigure}{.3\textwidth}
          \centering
          \includegraphics[width=\textwidth]{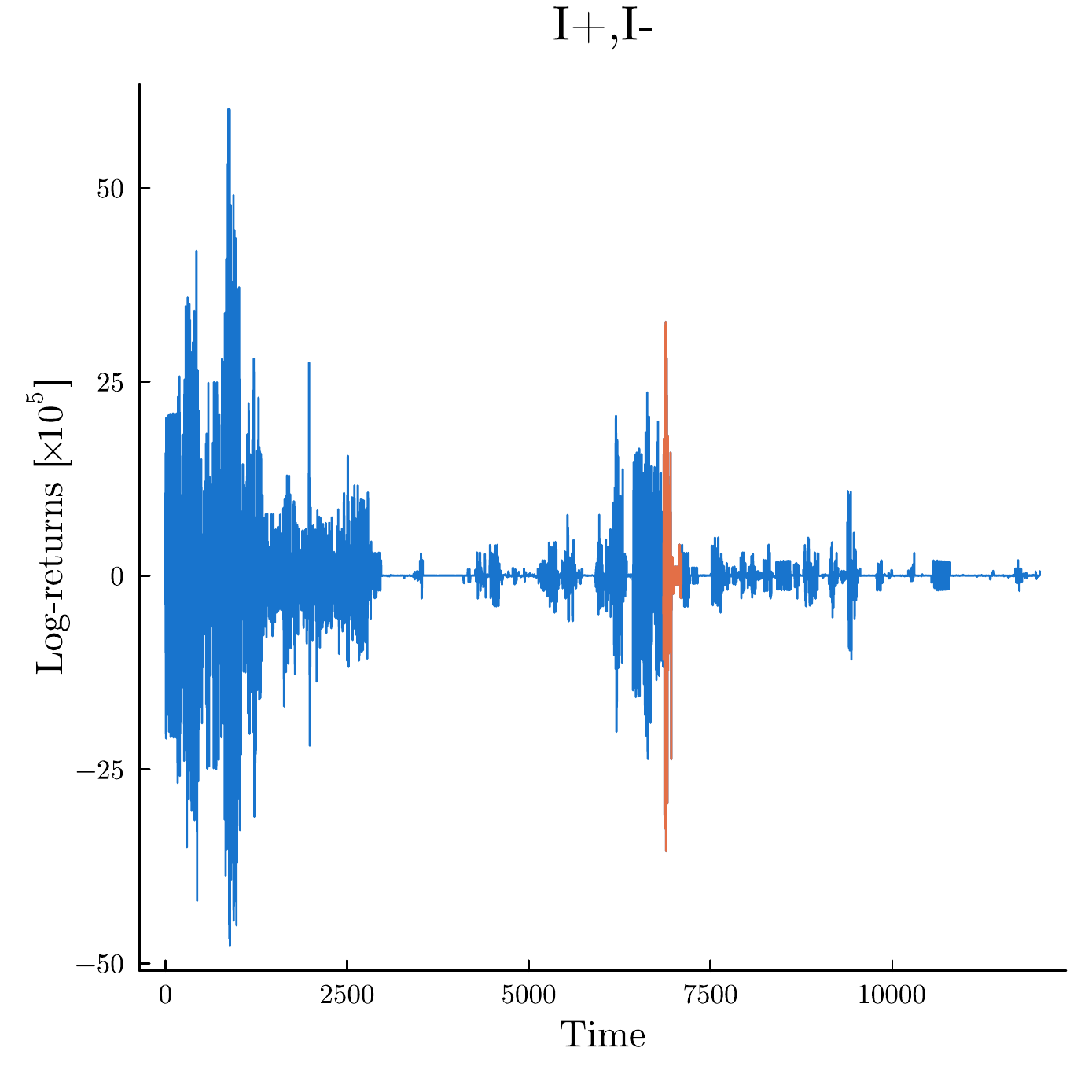}
          \caption{Case 5: micro-price fluctuations}
          \label{fig:time-series-plots-e}
        \end{subfigure}\\
        \begin{subfigure}{.3\textwidth}
          \centering
            \includegraphics[width=\textwidth]{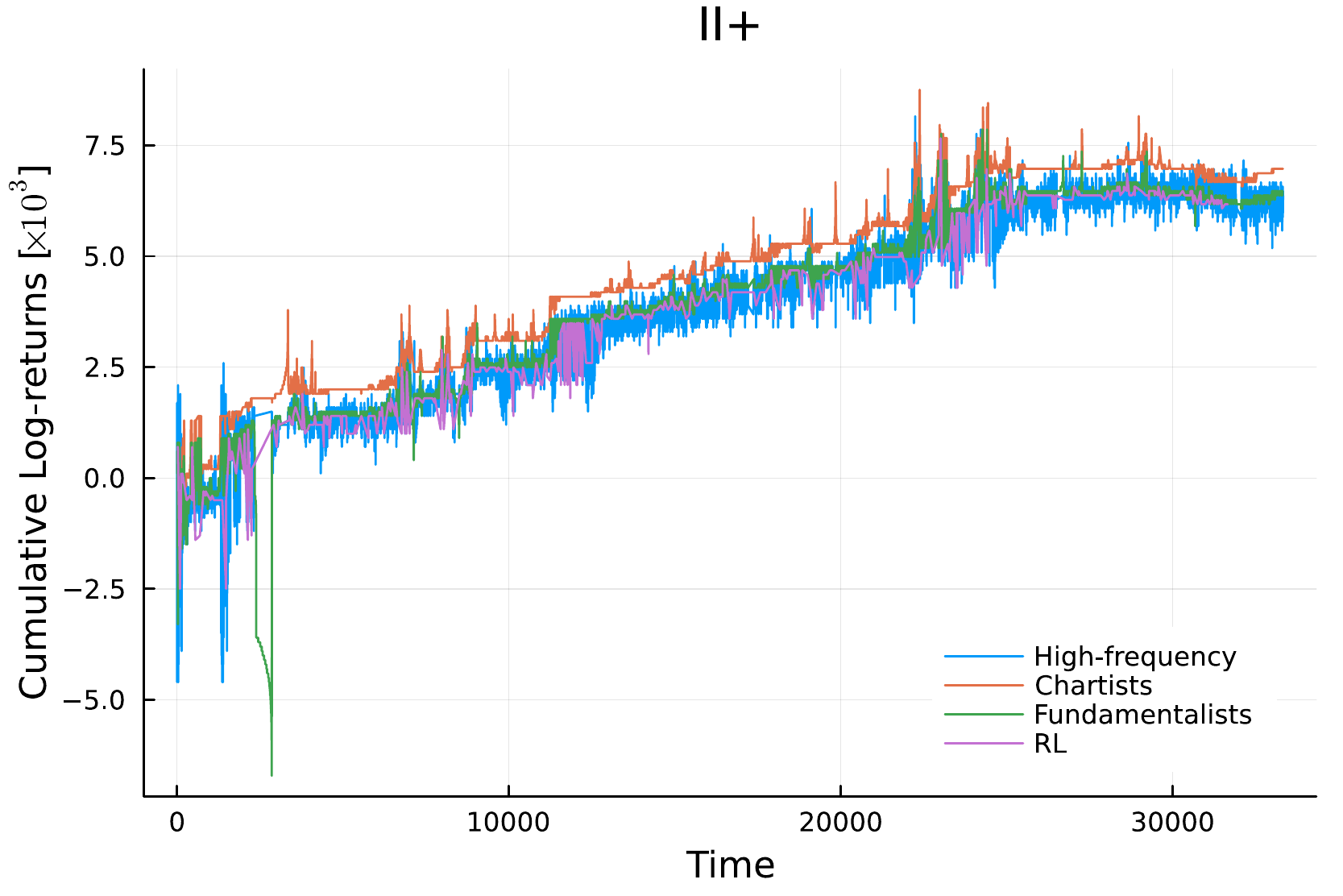}
              \caption{Case 6: Agent profitability}
              \label{fig:return-g}
        \end{subfigure}%
        \begin{subfigure}{.3\textwidth}
          \centering
          \includegraphics[width=\textwidth]{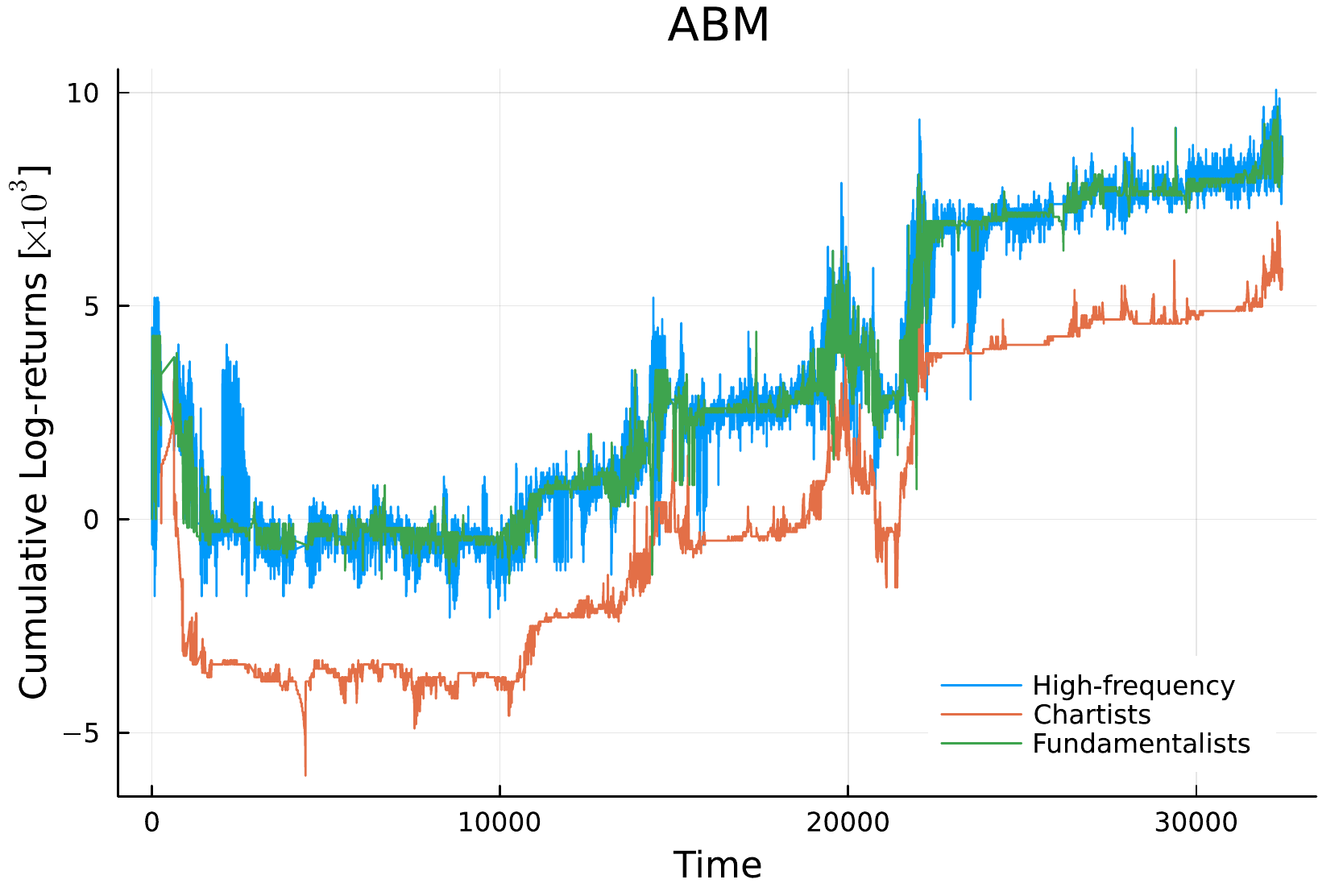}
          \caption{Case 0: Agent profitability}
          \label{fig:return-h}
        \end{subfigure}
                \begin{subfigure}{.3\textwidth}
          \centering
          \includegraphics[width=\textwidth]{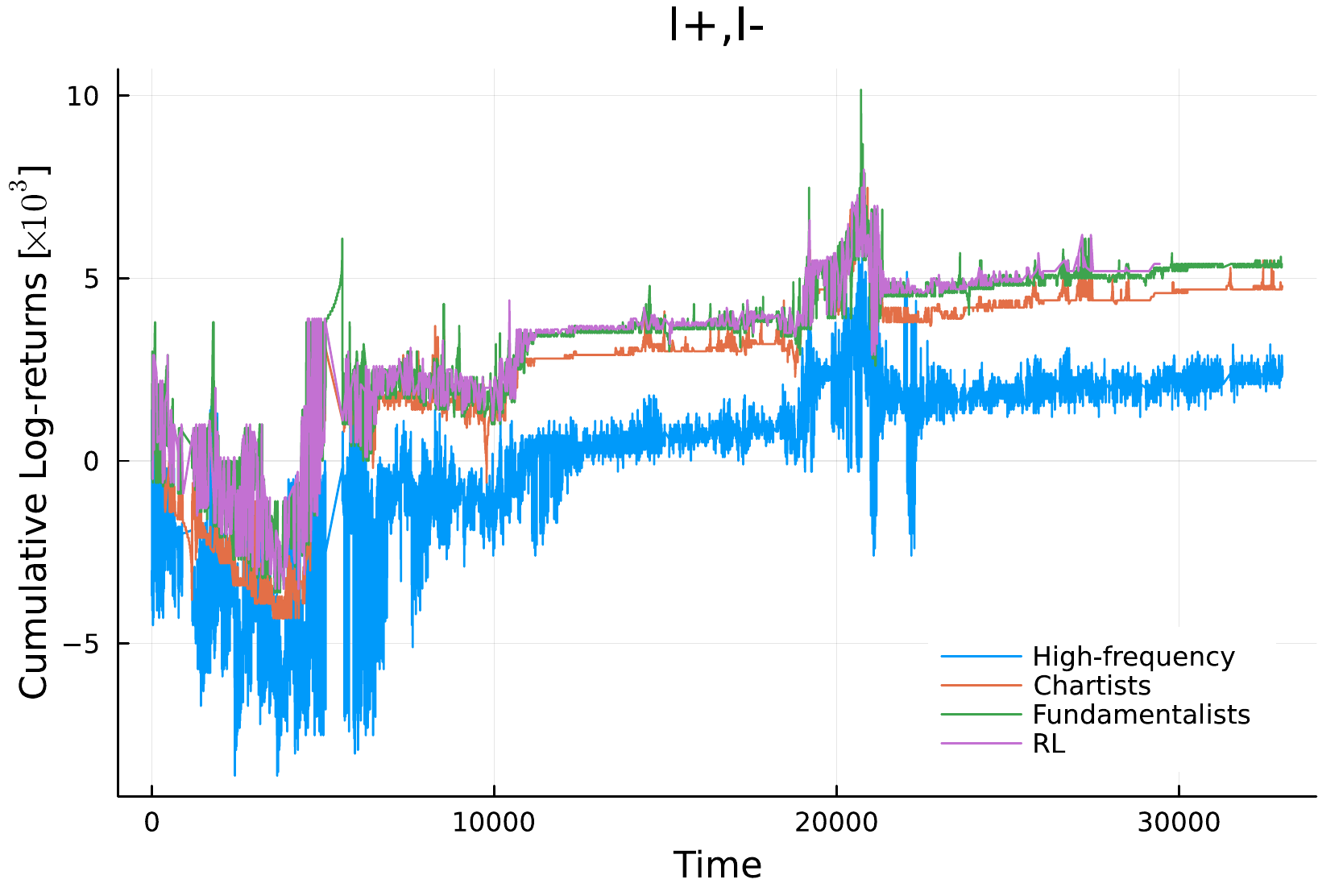}
          \caption{Case 5: Agent profitability}
          \label{fig:return-i}
        \end{subfigure}%
        \caption{The phase space reconstruction plots using a 2-dimensional embedding, and a delay times $\tau$ = 10, 10, and 6, respectively, for figures \ref{fig:phase-space-plots-a}, \ref{fig:phase-space-plots-b} and \ref{fig:phase-space-plots-c}. Segments, each of 250 points, are highlighted in red and show the dynamics associated with large micro-price movements, and these associated time-series segments are given in the second row of figures \ref{fig:time-series-plots-d}, \ref{fig:time-series-plots-e} and \ref{fig:time-series-plots-f}. The lower row of plots are equivalent to ``population" plots for the four different agent classes, the liquidity providers, the two classes of minimally intelligent liquidity takers, and the optimal execution agents where figures \ref{fig:return-g}, \ref{fig:return-h} and \ref{fig:return-i} plot the total running profit of the four agent classes. The plots are provided for Case 6, Case 0 and Case 5 as described in Table \ref{tab:RLagentscombinations}.}
        \label{fig:phase-space-plots}
\end{figure*}

We estimate the correlation dimensions using the method of \citet{GrassbergerProcaccia1983}. This gives us a reasonable bound on the required phase space from the micro-price data. The resulting correlation dimension will be used as a proxy for the representation dimension of a particular model configuration {\it i.e} as a measure of the model configuration's relative complexity. We select the correlation time in machine-time by selecting the first minimum in the micro-price autocorrelations \cite{Henryetal2005Chapter1}. 

\begin{figure*}[!h]
\begin{subfigure}[b]{0.48\textwidth}
    \includegraphics[width=1.0\textwidth]{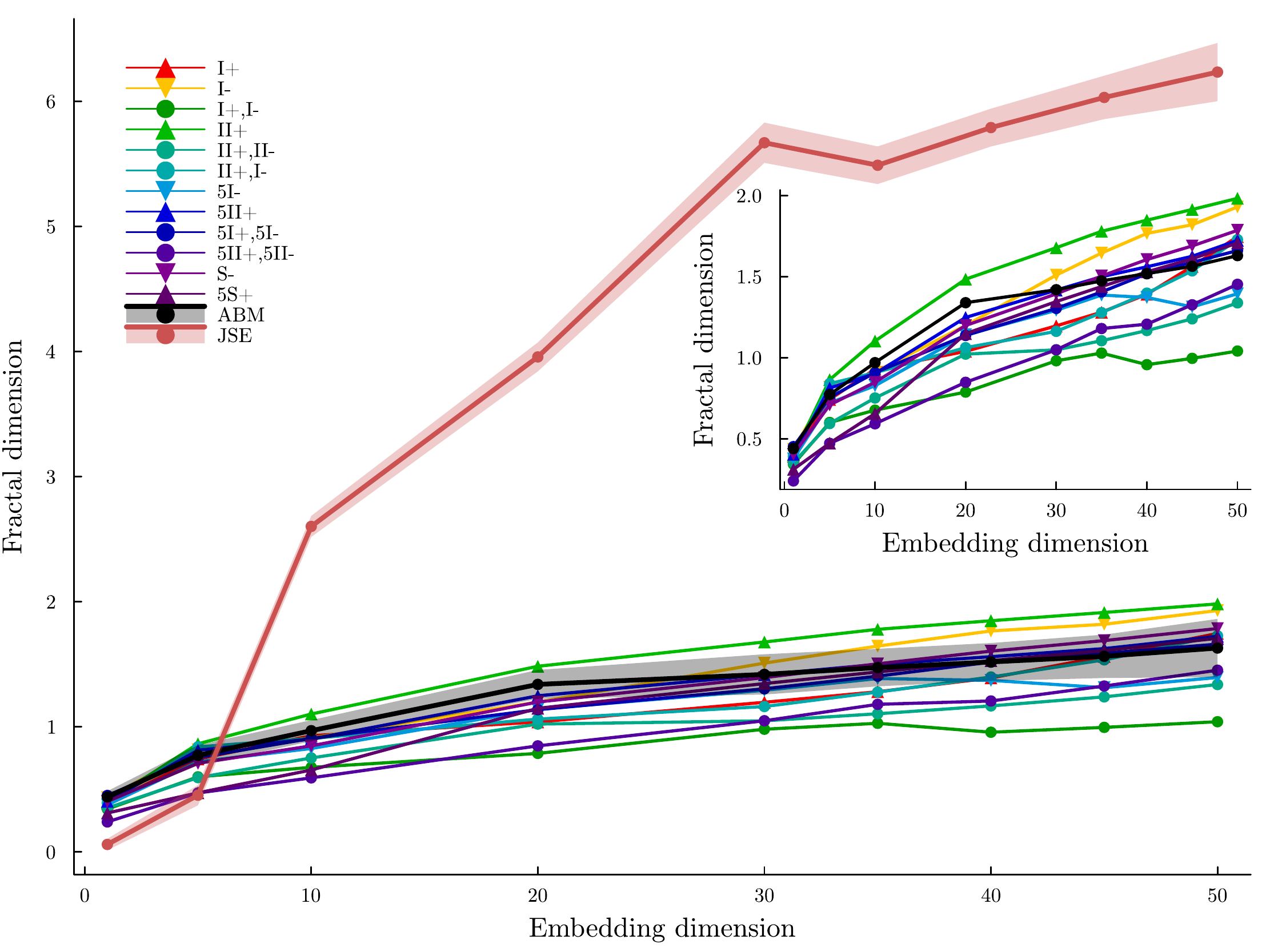}
    \caption{The fractal (correlation) dimension as a function of the embedding dimension found using the Grassberger-Procaccio algorithm \cite{GrassbergerProcaccia1983}. }
    \label{fig:embeddingdimensions}
\end{subfigure}
\begin{subtable}[b]{0.48\textwidth}
    \begin{tabular}{lcr|llll}
    \toprule
    Case & Types & $\Delta D$ & \multicolumn{2}{c}{\# RL} &  Exp. & $\overline{ \Delta D}$ \\
      \hline
      6   &  II$^+$ & 0.319 & 
     \parbox[t]{2mm}{\multirow{3}{*}{\rotatebox[origin=c]{90}{single}}} & \multirow{3}{*}{1} & \multirow{3}{*}{$X_0$} & \multirow{3}{*}{0.130} \\
      4   &  I$^-$  & 0.213 & & & &\\
      1   &  S$^-$ & 0.075 &\\
      3   &  I$^+$  & -0.086 & & & &\\
      \hline
      9   &  5II$^+$ & 0.044 & \parbox[t]{2mm}{\multirow{7}{*}{\rotatebox[origin=c]{90}{many agents}}}  & \multirow{2}{*}{5} & \multirow{2}{*}{$X_0$} & \multirow{2}{*}{-0.040} \\
      2   &  5S$^+$  & 0.005 & & & &\\
      10   &  5I$^-$  & -0.170 & & & &\\
      11   &  5I$^+$, 5I$^-$ & -0.026 & & \multirow{2}{*}{10} & \multirow{2}{*}{0} & \multirow{2}{*}{-0.152} \\
      12  & 5II$^+$, 5II$^-$ & -0.278 & & & &\\
      5   & II$^+$,I$^-$ & -0.101 & & \multirow{3}{*}{2} & \multirow{3}{*}{0} & \multirow{3}{*}{-0.321} \\
      7   & II$^+$,II$^-$ & -0.341 & & & &\\
      8   &  I$^+$,I$^-$  & -0.520 & & & &\\
    \bottomrule
    \end{tabular}
    \caption{Comparing the complexity of different model configurations. The single agents tend to increase the difference relative to the training environment, and the many agent configurations reduce the difference.}
    \label{tab:dimensiondifferences}
\end{subtable}
\caption{Comparing the complexity of the different configurations. There is complexity that is missing when compared to the real-world data (red) relative to the model distortions around the environment (black) introduced by the inclusion of learning agents. On the left (in Figure \ref{fig:embeddingdimensions}), we have that triangles represent simulations with only either buy or sell side agents and the circles represent simulations with both buy and sell side agents, as well as the JSE and ABM.}
\label{fig:complexitydifferences}
\end{figure*}

From Table \ref{tab:dimensiondifferences}, we can consider the relative differences in the model complexity measured at the higher embedding dimensions where there is the slowest increase in the estimated fractal dimension (the right of Figure \ref{fig:embeddingdimensions}). In particular when focusing on the difference on the right of the inset of Figure \ref{fig:embeddingdimensions}. Here we find that when single learning agents are combined with ABM, they tend to increase the fractal dimension, and hence, the complexity of the model configuration. Adding multiple learning agents tends to decrease the fractal dimension\,---\,possibly because these tend to interact with each other and reduce the overall impact of adaption and learning in the combined market. However, it is important to realise that the overall total outstanding initial parent order volume has been kept the same, at $X_0$=6\% of ADV (see Table \ref{tab:RLagentscombinations}). This means that the combined outstanding orders, the volume exposure (Vol. Exp.) for the multi (many) learning agents cases are {\it balanced} for cases 5,7,8,11 and 12, where the size of the buying position is the same as the selling position, zero. This is not the case for cases 1, 2, 3, 4, 6, 9 and 10 where there is non-zero volume exposure; the volume exposure is {\it one-sided}. This suggests that increases in the fractal dimension may be explained in this model to be due to asymmetric liquidity demand. 

To provide context for the complexity measures we provide phase space reconstruction plots using a 2-dimensional embedding in Figure \ref{fig:phase-space-plots}. Here with delay times $\tau$ = 10, 10, and 6, respectively, for figures \ref{fig:phase-space-plots-a}, \ref{fig:phase-space-plots-b} and \ref{fig:phase-space-plots-c}. Segments, each of 250 points, are highlighted in red and show the dynamics associated with large micro-price movements, and these associated time-series segments are given in the second row of figures \ref{fig:time-series-plots-d}, \ref{fig:time-series-plots-e} and \ref{fig:time-series-plots-f}. We notice that there is non-random structure to the dynamics with indications of quasi-periodic orbits. The plots are noticeably distorted relative to each other.  With the single learning agent (Case 6) having the most extended phase-space, and the situation with both a buying and selling learning agent (Case 5) having the most concentrated dynamics. The lower row of plots are equivalent to ``population" plots for the four different agent classes, the liquidity providers, the two classes of minimally intelligent liquidity takers, and the optimal execution agents where figures \ref{fig:return-g}, \ref{fig:return-h} and \ref{fig:return-i} plot the total running profit of the four agent classes.


\section{Conclusion}\label{sec:conclusion}

Our expanded description of market ecology includes optimal execution agents, which are necessary to produce stylised facts associated with order flow and the cost of trading. We further argue that learning is necessary because markets fail when the volume traded by a given agent exceeds the carrying capacity of the environment. However, having many agents attempting to learn simultaneously can result in complex non-stationary market dynamics, preventing successful learning for any agent.

We explicitly exclude strategy switching between the fundamentalists and chartists as a model design decision in the environment in which learning and adaption take place. This is premised on the reality that in real financial markets traders, money managers and investors cannot easily change strategies, and do not in fact change strategies because of how both treasury functions in investment banks allocate risk capital to strategies and trading desk, as well as regulatory structure surrounding mandate design specification for asset management and market-making functions. Here, we only allow strategy die-out. However, the learning agents can exploit the relative die-out and emergence of particular strategies \cite{Dicksetal2024ABM}. 

The inclusion of execution agents to a minimally intelligent ABM introduces additional heterogeneity into the observed order flow and liquidity provision processes. How these processes, and consequently the stylised facts, change depends on the specification of the optimal execution agents. In particular, we find that: \textit{i)} persistence in order flow increases with the number of execution agents trading on a single side, \textit{ii)} the realised cost of trading decreases when agents can submit LOs, and \textit{iii)} increasing the complexity of trading agents introduces an additional source of variation into the price process. These findings suggest the necessity of including optimal execution agents in ABMs to recover empirical high-frequency stylised facts. 

Furthermore, we find that learning introduces further variation into the order flow and liquidity processes, as agents make state-based decisions using a decision rule that adapts over time. This is demonstrated by the decrease in the level and persistence of autocorrelations. Surprisingly, we did not find that learning decreased the average cost of trading as reflected by the realised price impact functions. However, we still find evidence that learning is feasible since execution agents increase their performance over successive training periods. Although learning agents added complexity to the market's dynamics, this was insufficient to recover the complexity observed in empirical data.

In spite of the necessity of incorporating execution agents for a realistic ABM framework, neither execution agents or learning appears to be the dominant source of financial market complexity, at least when considered for a single stock in isolation from the broader market. Thus, as future work, we think it is worthwhile to consider the interaction of at least two markets, and investigate the emergence of correlations as the driver of the missing financial market complexity. This would allow for the investigation of a new agent class: multi-asset portfolio optimising agents, and how they are situated within the existing market ecology. In short, the bulk of the nonlinear dynamics, and complexity, in the single stock setting, seems to arise from the minimally intelligent agent dynamics, and it is this dynamic that provides the opportunity for learning.  
 
\section*{Code and Data Availability}

\section*{Acknowledgements}

We thank Ivan Jericevich for support and advice with respect to the matching engine and hybrid agent-based model implementation. We thank colleagues in the broader research community for helpful feedback and criticism. 

\section*{Author contributions statement}

T.G. and M.D. conceived the experiments and models,  M.D. implemented and conducted the experiments, A.P implemented and conducted the analysis, M.D, A.P. and T.G. analysed the results. All authors reviewed the manuscript. 

\section*{Competing interests} There are no competing interest. 

\bibliographystyle{elsarticle-harv}
\bibliography{MDAPTG-MALABM}

\appendix

\setcounter{table}{0}


\end{document}